\documentclass[reprint, prd, superscriptaddress, tightenlines, longbibliography, nofootinbib, eqsecnum, amsfonts, amsmath, floatfix, notitlepage,numbers,sort&compress,super]{revtex4-2}

\pdfoutput=1
\usepackage{textcomp}
\usepackage{bm,color}
\usepackage{verbatim}
\usepackage{amsmath}
\usepackage{ulem}
\usepackage[dvipsnames]{xcolor}
\usepackage{xcolor,colortbl}
\definecolor{linkcolor}{rgb}{0.6,0,0}
\definecolor{citecolor}{rgb}{0,0,0.75}
\definecolor{urlcolor}{rgb}{0.12,0.46,0.7}
\usepackage[breaklinks=true, colorlinks, urlcolor=urlcolor, linkcolor=linkcolor,citecolor=citecolor,pdfencoding=auto]{hyperref}
\usepackage{booktabs}
\usepackage{bigstrut}
\usepackage{xspace}
\usepackage{multirow}
\usepackage{amsmath}
\usepackage{graphicx}
\usepackage{threeparttable}
\usepackage{subfigure}
\usepackage{enumitem}

\newcommand{\nver}{\hat{\mathbf{n}}}

\newcommand{\vanish}[1]{}

\newcommand{\planck}{\textit{Planck}\xspace}

\def\aap{Astron.\ Astrophys.}

\def\apj{Astrophys.\ J.}

\def\jcap{J.\ Cosmol.\ Astropart.\ Phys.}

\def\mnras{Mon.\ Not.\ R.\ Astron.\ Soc.}

\def\prd{Phys.\ Rev.\ D}

\def\prl{Phys.\ Rev.\ Lett.}

\newcommand{\beq}{\begin{equation}}
\newcommand{\eeq}{\end{equation}}
\newcommand{\bea}{\begin{eqnarray}}
\newcommand{\eea}{\end{eqnarray}}

\begin{document}


\title{CMB-S4: Foreground-Cleaning Pipeline Comparison for Measuring Primordial Gravitational Waves}

\author{Federico Bianchini}
\affiliation{Kavli Institute for Particle Astrophysics and Cosmology, Stanford University, 452 Lomita Mall, Stanford, CA, 94305, USA}
\affiliation{Department of Physics, Stanford University, 382 Via Pueblo Mall, Stanford, CA, 94305, USA}
\affiliation{SLAC National Accelerator Laboratory, 2575 Sand Hill Road, Menlo Park, CA, 94025, USA}

\author{Dominic Beck}
\affiliation{Kavli Institute for Particle Astrophysics and Cosmology, Stanford University, 452 Lomita Mall, Stanford, CA, 94305, USA}
\affiliation{Department of Physics, Stanford University, 382 Via Pueblo Mall, Stanford, CA, 94305, USA}
\affiliation{SLAC National Accelerator Laboratory, 2575 Sand Hill Road, Menlo Park, CA, 94025, USA}

\author{W.L. Kimmy Wu}
\affiliation{Kavli Institute for Particle Astrophysics and Cosmology, Stanford University, 452 Lomita Mall, Stanford, CA, 94305, USA}
\affiliation{SLAC National Accelerator Laboratory, 2575 Sand Hill Road, Menlo Park, CA, 94025, USA}

\author{Zeeshan Ahmed}
\affiliation{Kavli Institute for Particle Astrophysics and Cosmology, Stanford University, 452 Lomita Mall, Stanford, CA, 94305, USA}
\affiliation{SLAC National Accelerator Laboratory, 2575 Sand Hill Road, Menlo Park, CA, 94025, USA}

\author{Sebastian Belkner}
\affiliation{Université de Genève, Département de Physique Théorique et CAP, 24 Quai Ansermet, CH-1211 Genève 4, Switzerland}

\author{Julien Carron}
\affiliation{Université de Genève, Département de Physique Théorique et CAP, 24 Quai Ansermet, CH-1211 Genève 4, Switzerland}

\author{Brandon S. Hensley}
\affiliation{Jet Propulsion Laboratory, California Institute of Technology, 4800 Oak Grove Drive, Pasadena, CA 91109, USA}

\author{Clement L. Pryke}
\affiliation{Minnesota Institute for Astrophysics, University of Minnesota, Minneapolis, MN 55455, USA}
\affiliation{School of Physics and Astronomy, University of Minnesota, Minneapolis, MN 55455, USA}

\author{Caterina Umilt\`a}
\affiliation{ Department of Physics, University of Cincinnati, Cincinnati, OH 45221, USA}
\affiliation{Department of Physics, University of Illinois at Urbana-Champaign, Urbana, IL 61801, USA}

\collaboration{CMB-S4 Collaboration}

\date{\today}

\begin{abstract}
We compare multiple foreground-cleaning pipelines for estimating the tensor-to-scalar ratio, $r$, using simulated maps of the planned CMB-S4 experiment within the context of the South Pole Deep Patch. 
To evaluate robustness, we analyze bias and uncertainty on $r$ across various foreground suites using map-based simulations. 
The foreground-cleaning methods include: a parametric maximum likelihood approach applied to auto- and cross-power spectra between frequency maps; a map-based parametric maximum-likelihood method; and a harmonic-space internal linear combination using frequency maps.
We summarize the conceptual basis of each method to highlight their similarities and differences.
To better probe the impact of foreground residuals, we implement an iterative internal delensing step, leveraging a map-based pipeline to generate a lensing $B$-mode template from the Large Aperture Telescope frequency maps.
Our results show that the performance of the three approaches is comparable for simple and intermediate-complexity foregrounds, with $\sigma(r)$ ranging from 3 to 5 $\times 10^{-4}$. 
However, biases at the $1-2\sigma$ level appear when analyzing more complex forms of foreground emission. 
By extending the baseline pipelines to marginalize over foreground residuals, we demonstrate that contamination can be reduced to within statistical uncertainties, albeit with a pipeline-dependent impact on $\sigma(r)$, which translates to a detection significance between 2 and 4$\sigma$ for an input value of $r = 0.003$. 
These findings suggest varying levels of maturity among the tested pipelines, with the auto- and cross-spectra-based approach demonstrating the best stability and overall performance.
Moreover, given the extremely low noise levels, mutual validation of independent foreground-cleaning pipelines is essential to ensure the robustness of any potential detection.
\end{abstract}

\maketitle


\section{Introduction \label{sec:intro}}

The cosmic microwave background (CMB) provides a unique window into the early Universe and inflation, e.g. Refs.~\citep{starobinsky80,guth81,sato81,linde82}. 
In particular, the parity-odd $B$ modes of CMB polarization are expected to be in part produced by primordial gravitational waves (PGWs) generically generated during inflation, e.g. Refs.~\citep{polnarev85,Seljak97,Kamionkowski97}.
The tensor-to-scalar ratio $r$ describes the amplitude of primordial gravitational waves relative to the scalar perturbations. 
Measuring $r$ discriminates between different inflation models and probes energy scales beyond those accessible by particle colliders, e.g. Refs.~\citep{Kamionkowski_2016,achucarro2022inflation}. 
Due to the profound implications for fundamental physics in detecting gravitational waves from inflation, numerous experiments have been designed to probe the large-scale $B$ modes of the CMB, where the primordial gravitational wave $B$-mode signal is expected to peak, e.g. Ref.~\cite{chang2022snowmass2021}.
Notable among these are CLASS \citep{class2014}, SPIDER \citep{spider2022}, the Simons Observatory \citep{SO,wolz24}, South Pole Observatory, CMB-S4 \citep{s4forecast,s4sciencebook}, AliCPT \citep{alicpt}, and LiteBIRD \citep{litebird}. \\

The highest sensitivity observations and strongest constraints come from the measurements by the BICEP/Keck Collaboration~\citep[BK18,][]{bk18},  $r_{0.05} < 0.036$ at 95\% credibility level (95\% C.L.), reaching $r < 0.032$ when combined with the latest analysis of the \planck data (PR4) and Baryonic Acoustic Oscillations data~\cite{Tristram22}.\footnote{The bound from Ref.~\citep{Tristram22} relaxes to $r < 0.038$ when a conditioned \planck low-$\ell$EB covariance matrix is used \citep{Beck2022,Campeti_2022,de_Belsunce_2022}.}
The forthcoming ground-based CMB-S4 experiment is planned to test inflationary models by aiming to either place an upper limit $r \leq 0.001$ at 95\% C.L. if $r = 0$, or detect $r$ at $\ge 5\sigma$ if $r > 0.003$ \citep{s4forecast,s4sciencebook}. 
A detection of $r > 0.003$ would imply inflationary physics near the energy scale of grand unified theories. 
Many inflationary models that accommodate the observed tilt in the scalar perturbation spectrum ($n_{\rm s} < 1$) and predict a characteristic field excursion larger than the Planck mass anticipate tensor-to-scalar ratios exceeding $0.001$, with well-motivated subclasses predicting $r > 0.003$. 
This sets the stage for a potentially transformative period in early-Universe physics, where popular models such as those of Starobinsky $R^2$ \citep{starobinski79} and Higgs \citep{Bezrukov_2008} inflation can be conclusively tested.\\

Probing these primordial $B$ modes poses significant instrumental and observational challenges.
The expected signal is faint and its amplitude unknown, necessitating high sensitivity observations and meticulous control of instrumental systematics that could mimic $B$-mode power \citep[e.g.,][]{odea07,Shimon_2008,Yadav10,soliman18, Verges2021}. 
Additionally, primordial $B$ modes are obscured by lensing-induced $B$ modes arising from the large-scale structure \citep[e.g.,][]{Knox_2002,Kesden_2002,Seljak_2004}. 
A major challenge in the search for PGWs with ultra-low-noise experiments is the contamination by astrophysical $B$-mode polarization from our own Galaxy. 
Dominated by dust and synchrotron emission at high ($\nu \gtrsim 60$ GHz) and low ($\nu \lesssim 60$ GHz) frequencies respectively, much of our current understanding of Galactic foregrounds at millimeter-wavelengths stems from \planck data \citep[e.g.,][]{planck18_diffuse_compsep,Krachmalnicoff18}. 
Effectively separating these polarized Galactic foregrounds from the primordial signal is crucial for the success of CMB-S4.\\

Disentangling foregrounds from the primordial $B$-mode signal is achieved through various component-separation methods, each differing in their data modeling approaches and assumptions about the components to be separated. These methods are broadly categorized into parametric and non-parametric techniques, as well as map-based and power-spectrum-based approaches, depending on the operational space.
Non-parametric methods, often referred to as ``blind," typically rely on different properties of the signals, such as statistical independence or sparsity, and a known CMB frequency scaling to separate the different components. Examples include the Internal Linear Combination (ILC) and its variants \citep{Tegmark1996,Bennett_2003,Delabrouille_2008,remazeilles10}, Fast Independent Component Analysis (FastICA) \citep{Maino_2002}, Local Generalized Morphological Component Analysis (L-GMCA) \citep{Bobin:2007hf,Bobin:2012hq}, and the Minimally Informed CMB MAp foreground Cleaning method (MICMAC) \citep{Leloup2023,Morshed2024}.
Parametric methods, on the other hand, are based on phenomenological modeling of the sky components. 
They involve parameterizing the spectral dependence of the emission law for a given foreground component and estimating the free parameters from the data. While some parametric methods produce maps of the cosmological signal by marginalizing over the foreground parameters, others operate directly on the power spectra.
Notable examples of non-blind methods include FGBuster \citep{Stompor2009,Errard_2011}, the Bayesian CMB Gibbs sampler Commander \citep{Eriksen2006,Eriksen08}, B-SeCRET \citep{de_la_Hoz_2022}, and the Maximum Entropy Method (MEM) \citep{Stolyarov:2004xp}.
Additionally, there are intermediate methods that use prior information about the possible structure of the mixing operator, allowing for adjustable levels of blindness. Examples of such semi-blind methods are Spectral Matching ICA (SMICA) \citep{cardoso2008component} and Correlated Component Analysis (CCA) \citep{Bedini_2005}. Recently, hybrid methods that combine features from both map-based and $C_\ell$-based techniques have been proposed, e.g. Ref.~\citep{Azzoni_2023}.
Finally, recent multi-clustering techniques, currently applied to all-sky surveys, leverage existing data on diffuse foregrounds to optimize component separation. 
These methods identify distinct sky regions (``clusters") where the signal characteristics align with the assumptions of specific ILC-based \citep{Carones2023} or parametric \citep{Puglisi:2021hqe} foreground-cleaning techniques, thus improving their performance.\\

In this paper, we explore a range of choices within these categories to assess how each component-separation method performs under the foreground simulations provided for the CMB-S4 South Pole Deep Patch \cite{s4forecast}. 
While recent shifts in CMB-S4 strategic priorities and considerations have led to a reevaluation of the planned survey configuration to move away from the South Pole, the comparisons we present here remains valid for the considered cases.
Specifically, by analyzing performance across multiple component-separation methods at these extremely low noise levels, the biases from incorrectly modeling foregrounds and lensing become more significant.
This study thus provides relevant insights for future $r$ analyses, independent of survey design.\footnote{For a study focused on foreground-cleaning for $r$ for a survey from the Chilean site, the Simons Observatory, please see Ref.~\citep{wolz24}.}\\

The paper is structured as follows. In Sec.~\ref{sec:scope}, we present the scope of the paper, followed by a review of the theoretical foundations of each foreground-cleaning method in Sec.~\ref{sec:methods}, where we highlight their similarities and differences. 
Section~\ref{sec:skymodel} introduces the simulation suites, sky components, and noise models used in the analysis, while Sec.~\ref{sec:delensing} briefly covers the construction of the lensing template. 
The likelihood analysis framework is detailed in Sec.~\ref{sec:likelihood}, and the comparison of results from the different component-separation methods is discussed in Sec.~\ref{sec:results}.
We conclude in Sec.~\ref{sec:conclusions}.

\section{Scope}\label{sec:scope}
One of the main science goals of CMB-S4 is to measure the tensor-to-scalar ratio, $r$. 
The pertinent data-analysis task is to characterize the posterior density
\begin{equation}
\mathcal{P}(r|\mathbf{d}),
\end{equation}
where $\mathbf{d}$ is the data vector from CMB-S4 observations. 
The latter can be given most generally as time-ordered data, but for practical reasons, however, we usually evaluate this posterior after the data have undergone a series of data-reduction and processing steps. These steps can typically include, for Small Aperture Telescopes (SATs) data, time-domain filtering, followed by map making, instrumental systematic mitigation, component separation, and delensing, as well as power-spectrum estimation. 
Any statistical uncertainty has to be propagated to the final posterior evaluation through the covariance matrix.
We limit the scope of this paper to pipelines starting from maps at each of the $N_{\rm freq}=9$ planned observing frequencies of CMB-S4 SAT in the range between 20 and 270~GHz (see Tab.~\ref{tab:s4_spec}) free of instrumental systematics and investigate different approaches to component separation.\\

We consider different choices for when the maps and/or $r$ estimates are cleaned of foregrounds: specifically, whether foregrounds are removed before or after the power-spectrum-estimation step. We refer to the former as \textit{map-based} component separation and the latter as \textit{power-spectrum-based} (or $C_\ell$-based) component separation. This distinction affects the weighting of the input maps, which may cause the methods to respond to slightly different noise modes.\\

To fully leverage multi-frequency observations, we require a model to distinguish the CMB from foregrounds. This model may be either parametric (assuming specific functional forms for the frequency and/or spatial dependence of each component) or non-parametric (making fewer assumptions about these functional forms). For this study, we focus on three approaches: 1) a parametric power-spectrum-based method; 2) a parametric map-based method; and 3) a non-parametric map-based method (sometimes referred to as ILC in the literature). These choices reflect a range of possible strategies relevant to CMB-S4, representing different trade offs between model assumptions and flexibility in handling complex foregrounds.\\

In addition to foreground cleaning, we ensure that the $B$-mode power spectrum estimation is free of $E$-to-$B$ leakage and construct a parameter likelihood that incorporates delensing. The delensing method used in this work involves creating a lensing $B$-mode template, characterizing the cross-correlation between this template and the true $B$-mode signal, and accounting for the template’s noise. We note that neither delensing nor foreground cleaning has been applied to data at these low noise levels before. The successful recovery of an unbiased $r$ value from simulations demonstrates the overall adequacy of our modeling. Nevertheless, we highlight a potential issue with the lensing template's noise spectrum in Sec.~\ref{sec:delensing}.
\\

\begin{figure*}
    \centering
    \includegraphics[width=\linewidth]{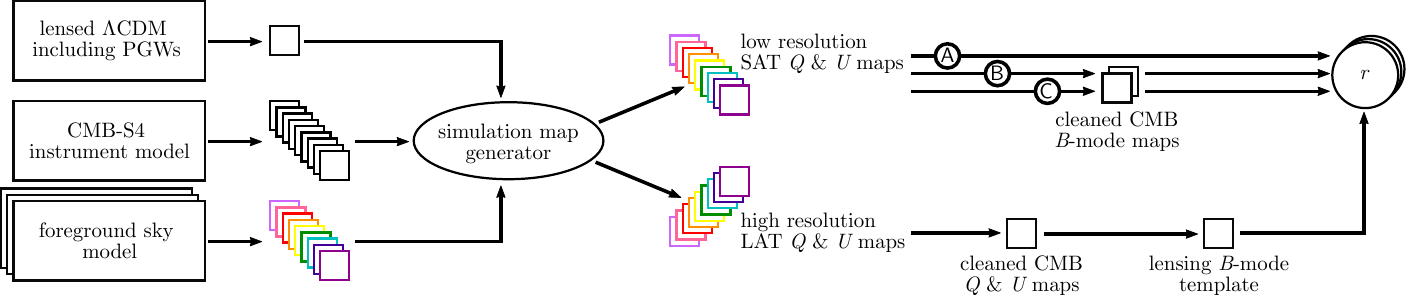}
    \caption{Flowchart of the simulation generation and foreground cleaning process. The simulation generation is described in Sec.~\ref{sec:skymodel} and includes simulations of the CMB, CMB-S4 noise at each frequency as well as multifrequency simulations of various foreground models. These simulated maps are fed into three different component separation pipelines (labeled A, B, and C) as well as a realistic lensing $B$-mode reconstruction pipeline. The outputs of these pipelines are passed to a likelihood to estimate the bias and uncertainty on $r$.}
    \label{fig:skymodelflowchart}
\end{figure*}

The main result of this paper is the validation of foreground-cleaning pipelines outlined in Fig.~\ref{fig:skymodelflowchart}. 
Each element of the pipeline will be discussed in the following sections. 
Using the CMB-S4 noise model and various foreground models, we simulate maps of the CMB-S4 SATs and Large Aperture Telescope (LAT) across multiple frequencies.
We apply three different foreground-cleaning pipelines to the SATs and a lensing $B$-mode template pipeline to the LAT. 
These pipelines are validated and their performance at the $r$-inference level is compared using a set of three different foreground suites. 
Subsequently, we test the pipelines on a separate set of simulations with different foregrounds, treating them as if they were real data. This demonstrates how mitigation strategies sufficient for one set of foregrounds can introduce biases when applied to another.
By including both ILC and parametric methods, our analysis provides a complete assessment of foreground cleaning using current component-separation techniques.

\section{Component Separation Methods \label{sec:methods}}
In subsections A-C, we first lay out the statistical foundation of each foreground-cleaning method and then specify the implementation of each method for this work (D).

\subsection{Likelihood}
In this paper, we assume our dataset to be in the form of maps of $Q$ and $U$ polarization Stokes parameters measured in each of CMB-S4's frequency bands (see Tab.~\ref{tab:s4_spec} for the technical specifications for a CMB-S4 SAT).
We base our inference on the linear data model
\begin{equation}\label{eq:data_model}
    \mathbf{d}=\mathbf{A}\mathbf{s}+\mathbf{n},
\end{equation}
where $\mathbf{s}$ are the separate sky components (e.g. CMB, dust) we want to solve for with the map-based pipelines, $\mathbf{A}$ is a linear operator, and $\mathbf{n}$ is the noise.
We decompose this linear operator and write it as
\begin{equation}
    \mathbf{A}\equiv\mathbf{R}\mathbf{B}\mathbf{F},
\end{equation}
where $\mathbf{R}$ is a filtering operator that describes some filtering applied to the input maps, which may be different between different methods, $\mathbf{B}$ is the beam-convolution operator, which convolves each frequency map with an appropriate beam, and $\mathbf{F}$ is the mixing matrix, which describes the mixing of sky components, $\mathbf{s}$, into observed frequency maps. 
The modeling can be performed under different assumptions and in different basis spaces, which can lead to crucial differences between inference methods.
Generally this requires the assumption of a statistical distribution of the noise, which we assume to be Gaussian, entirely described by the noise covariance
\begin{equation}
\mathbf{N}\equiv\left\langle\mathbf{n}\mathbf{n}^T\right\rangle.
\end{equation}

All our inference methods are based on the same likelihood function which encodes the probability of the observed data given some cosmological parameters (in particular $r$) and potential nuisance parameters, $\theta$,
\begin{equation}
    -2\log P(\mathbf{d}|\mathbf{s},\theta) = (\mathbf{d}-\mathbf{A}\mathbf{s})^T \mathbf{N}^{-1} (\mathbf{d}-\mathbf{A}\mathbf{s})
    \label{eq:masterlikelihood}.
\end{equation}
These nuisance parameters can for example determine the foreground model or describe some instrumental systematic effect, which can enter the likelihood in both the linear operator $\mathbf{A}$ or as priors on the sky components $\mathbf{s}$.

\subsection{Prior Choices}
Different prior choices lead to different foreground cleaning implementations (see also Ref.~\citep{Leloup2023}).
Specifically, with flat priors on the sky signal $\mathbf{s}$, we derive the framework for the map-based parametric method.
With Gaussian priors on $\mathbf{s}$, we derive the power-spectrum-based parametric as well as the map-based non-parametric method.

\subsubsection{Flat prior for the parametric map-based method}\label{sec:flatpriors}

Given the likelihood in Eq. \ref{eq:masterlikelihood}, flat priors on the sky signal $\mathbf{s}$ and an estimate of the noise covariance $\hat{\mathbf{N}}$, we arrive at the posterior distribution
\begin{equation}
    -2\log P(\mathbf{s},\theta|\mathbf{d}) = (\mathbf{d}-\mathbf{A}\mathbf{s})^T \hat{\mathbf{N}}^{-1} (\mathbf{d}-\mathbf{A}\mathbf{s}),
    \label{eq:flatposterior}
\end{equation}
with the maximum a-posteriori (MAP) solution for the sky signal
\begin{equation}
    \hat{\mathbf{s}} = \left( \mathbf{A}^T \hat{\mathbf{N}}^{-1} \mathbf{A} \right)^{-1} \mathbf{A}^T \hat{\mathbf{N}}^{-1} \mathbf{d}.
    \label{eq:mapskysignal}
\end{equation}
Given that both the content of the $\mathbf{A}$ operator and the sky signal $\mathbf{s}$ are unknown, we can make use of the fact that one can marginalize analytically over the latter in order to obtain a posterior distribution on the foreground model. This results in the posterior for the spectral parameters $\theta$ of the foreground model encoded as part of $\mathbf{A}$ in the mixing matrix $\mathbf{F}$,
\begin{align}
    -2\log P(\theta|\mathbf{d}) = &
    - \left(\mathbf{A}^T \mathbf{N}^{-1} \mathbf{d}\right)^T \left( \mathbf{A}^T \mathbf{N}^{-1} \mathbf{A} \right)^{-1} \left(\mathbf{A}^T \mathbf{N}^{-1} \mathbf{d}\right) - \notag\\ &- \log\det\left(\left( \mathbf{A}^T \mathbf{N}^{-1} \mathbf{A} \right)^{-1}\right).
\end{align}
Ref. \cite{Stompor2009} argues for the use of the spectral likelihood---the profile likelihood of Eq. \ref{eq:masterlikelihood} evaluated at the maximum likelihood solution given in Eq. \ref{eq:mapskysignal}---instead which yields unbiased parameter estimates 
\begin{align}
    -2\log P(\theta|\mathbf{d}) = &
    - \left(\mathbf{A}^T \mathbf{N}^{-1} \mathbf{d}\right)^T \left( \mathbf{A}^T \mathbf{N}^{-1} \mathbf{A} \right)^{-1} \left(\mathbf{A}^T \mathbf{N}^{-1} \mathbf{d}\right).
    \label{eq:spectrallikelihood}
\end{align}
Once maximized, the spectral parameters can be inserted in Eq.~\ref{eq:mapskysignal} to solve for $\hat{\mathbf{s}}$.

\subsubsection{Gaussian prior for the non-parametric map-based and parametric power-spectrum-based methods}

In addition to the spectral information about the sky components encoded in the mixing matrix $\mathbf{F}$, it is also possible to incorporate prior information on the spatial properties of any sky component.
It is straightforward to add a Gaussian prior on the sky signal, $\mathbf{s}$, by demanding it to be a Gaussian random field with covariance $\mathbf{S}$. The posterior now reads, 
\begin{align}
    -2\log P(r,\theta|\mathbf{d}) &=  (\mathbf{d}-\mathbf{A}\mathbf{s})^T \hat{\mathbf{N}}^{-1} (\mathbf{d}-\mathbf{A}\mathbf{s}) +\notag\\ &+ \mathbf{s}^T \mathbf{S}^{-1} \mathbf{s}  + \log\det \mathbf{S}.
\label{eqn:gausspriorpost}
\end{align}

Like above, we can write down the MAP solution for the sky signal $\mathbf{s}$
\begin{align}
    \hat{\mathbf{s}} &= \left( \mathbf{A}^T \hat{\mathbf{N}}^{-1} \mathbf{A} + \mathbf{S}^{-1}\right)^{-1} \mathbf{A}^T \hat{\mathbf{N}}^{-1} \mathbf{d} =\notag\\&= \mathbf{S}  \mathbf{A}^T   \left( \mathbf{N} + \mathbf{A} \mathbf{S} \mathbf{A}^T\right)^{-1}  \mathbf{d},
\end{align}
which is a Wiener filter.\\

In the literature \cite{Tegmark1996} one usually considers a modified estimator, the so-called ILC estimator, that is as close as possible to the MAP solution, but preserves power of the signal and is thus an unbiased estimator of $\mathbf{s}$, 
\begin{align}
    \hat{\mathbf{s}}_{\textrm{ILC}} &= \left( \mathbf{S} \mathbf{A}^T   \left( \mathbf{N} + \mathbf{A} \mathbf{S} \mathbf{A}^T\right)^{-1} \mathbf{A} \right)^{-1} \hat{\mathbf{s}} =\notag\\&=  \left( \mathbf{A}^T   \left( \mathbf{N} + \mathbf{A} \mathbf{S} \mathbf{A}^T\right)^{-1} \mathbf{A} \right)^{-1}  \mathbf{A}^T   \left( \mathbf{N} + \mathbf{A} \mathbf{S} \mathbf{A}^T\right)^{-1}  \mathbf{d} =\notag\\&=  \left( \mathbf{A}^T  \mathbf{C}^{-1} \mathbf{A} \right)^{-1}  \mathbf{A}^T   \mathbf{C}^{-1}  \mathbf{d}.
    \label{eq:ilcestimator}
\end{align}
A further convenience of this estimator is that it only depends on the total variance in the frequency maps $\mathbf{C}= \mathbf{N} + \mathbf{A} \mathbf{S} \mathbf{A}^T$, which might in practice be simpler to estimate than the noise-only covariance. \\

To arrive at the power-spectrum-based parametric solution, one analytically marginalizes the posterior in Eq.~\ref{eqn:gausspriorpost} over the sky signal to get,
\begin{align}
    -2\log P(r,\theta|\mathbf{d}) &= 
     \mathbf{d}^T \left(   \mathbf{N}+ \mathbf{A} \mathbf{S} \mathbf{A}^T \right)^{-1} \mathbf{d} +\notag\\&+\log\det\left( \mathbf{N} + \mathbf{A}\mathbf{S}\mathbf{A}^T \right).
\label{eqn:hl_post}
\end{align}
This can be rewritten as a likelihood in terms of bandpowers, and is commonly approximated as a likelihood quadratic in the observed bandpowers accounting for cut-sky effects with a simulation-based bandpower-by-bandpower covariance matrix~\citep[e.g. Hammimeche-Lewis, or HL, likelihood][]{Hamimeche2008}.\\

\subsection{Choices for the sky component model}

A model of the foregrounds can enter the posterior distributions above in the mixing matrix $\mathbf{F}$ and in the model for the covariance matrix $\mathbf{S}$.
Most generally, the matrix $\mathbf{F}_{p,i}$ has dimensions $N_\textrm{freq}\times N_\textrm{comp}$ for each pixel or multipole $p$ and Stokes parameter $i$. \\

Assuming CMB as the only sky component and  equal brightness in each frequency,  common choices for the construction of this matrix includes $N_\textrm{comp}=1$, such that $\mathbf{F}_{p,i}$ has the form
\begin{align}
    \mathbf{F}_{p,i}=(1\ ...\ 1)^T.
    \label{eq:1componentmixing}
\end{align}
Another option is to assume and reconstruct a Galactic dust and synchrotron component, respectively, such that $N_\textrm{comp}=3$. Again assuming equal brightness of the CMB in each frequency, that leaves us at $2\times N_\textrm{freq}$ free matrix elements. We will employ a parametric model for these matrix elements such that the mixing matrix only depends on the two free spectral parameters $\beta_{\rm d}$ and $\beta_{\rm s}$ 
\begin{align}
    \mathbf{F}_{p,i}=
    \begin{pmatrix}
        1           & ... & 1 \\
        f_{\rm d}^{\nu_1} & ... & f_{\rm d}^{\nu_N} \\
        f_{\rm s}^{\nu_1} & ... & f_{\rm s}^{\nu_N}
    \end{pmatrix}^T,
    \label{eq:3componentmixing}
\end{align}
where we assume a modified blackbody spectral energy distribution (SED) for dust and a power-law SED for synchrotron
\begin{align}
    f_{\rm d}^{\nu} &\propto \int d\nu R(\nu) \left(\frac{\nu}{\nu_\textrm{pivot}}\right)^{3+\beta_{\rm d}} \left(\exp\frac{h\nu}{k T_{\rm d}}-1\right)
    \label{eq:dustSED}\\
    f_{\rm s}^{\nu} &\propto \int d\nu R(\nu) \left(\frac{\nu}{\nu_\textrm{pivot}}\right)^{2+\beta_{\rm s}},
    \label{eq:syncSED}
\end{align}
with the dust temperature $T_{\rm d}$, fixed to $T_{\rm d}=19.6\ \textrm{K}$ throughout the paper, and the frequency bandpass $R(\nu)$.\footnote{In this analysis, we fix the dust temperature to $T_{\rm d}=19.6\ \textrm{K}$ because the available frequency coverage, particularly at higher frequencies, is insufficient to effectively constrain it as a free parameter.}
We assume pivot frequencies, $\nu_\textrm{pivot}$, of 353 GHz and 23 GHz for Galactic dust and synchrotron, respectively.
When integrating the SED and conversion factors over a given bandpass, we adopt the convention that our bandpasses describe the response as a function of frequency to a beam-filling source with uniform spectral radiance. See the Appendix of \citep{s4forecast} for more details on constructing these coefficients.

\subsection{Pipeline choices}

Following the previous sections we can build pipelines with varying assumptions on the spectral and spatial properties of the sky components. 

\subsubsection{Pipeline A (Parametric cross-$C_\ell$ based method)}\label{sec:crossCl}
Pipeline A follows very closely the parametric multi-frequency cross-spectral approach used by the BICEP/Keck collaboration \citep{bk15}, and as used in our previous CMB-S4 forecast paper \citep{s4forecast}.
In this pipeline we make parametric assumptions on both the spectral and spatial properties of the sky components. \\

This power-spectrum-based method involves fitting a multi-component sky model to the observed auto- and cross-frequency $B$-mode auto-power spectra $C_\ell^{\nu\nu'}$ computed between all frequency maps using the posterior in Eq.~\ref{eqn:hl_post} as the likelihood. \\

The model for $\mathbf{S}$ includes the fixed lensed $\Lambda$CDM and a PGW $BB$ theoretical power spectrum, as well as contributions from Galactic dust and synchrotron emissions parameterized as
\begin{align}
C_{\ell}^{\textrm{d},\nu_1 \nu_2} &=  A_{\rm{d}} \Delta_{\mathrm{d}}^{\prime} f_{\mathrm{d}}^{\nu_1} f_{\mathrm{d}}^{\nu_2}\left(\frac{\ell}{80}\right)^{\alpha_{\mathrm{d}}}, \\
C_{\ell}^{\textrm{s},\nu_1 \nu_2} &=  A_{\mathrm{s}} \Delta_{\mathrm{s}}^{\prime} f_{\mathrm{s}}^{\nu_1} f_{\mathrm{s}}^{\nu_2}\left(\frac{\ell}{80}\right)^{\alpha_{\mathrm{s}}}, \\
C_{\ell}^{\textrm{d}\times\textrm{s},\nu_1 \nu_2} &=  \epsilon \sqrt{A_{\mathrm{d}} A_{\mathrm{s}}} \left(f_{\mathrm{d}}^{\nu_1} f_{\mathrm{s}}^{\nu_2} + f_{\mathrm{s}}^{\nu_1} f_{\mathrm{d}}^{\nu_2}\right)\left(\frac{\ell}{80}\right)^{\left(\alpha_{\mathrm{d}}+\alpha_{\mathrm{s}}\right) / 2}.
\label{eqn:bk_fgmodel}
\end{align}
The parameters $A_{\rm d}$ and $A_{\rm s}$ characterize the dust and synchrotron power (in units of $\mu{\rm K}^2$) at scale $\ell = 80$.
The spatial fluctuations of both foregrounds are assumed to follow a power-law spectrum with slopes $\alpha_{\rm d}$ and $\alpha_{\rm s}$ for dust and synchrotron, respectively.
The amount of spatial correlation between dust and synchrotron is parameterized by $\epsilon$, which we assume to be scale-independent.
As such, the correlated component scales in $\ell$ with a slope given by the average between $\alpha_{\rm d}$ and $\alpha_{\rm s}$.
The parameters $\Delta_{\rm d}'$ and  $\Delta_{\rm s}'$ account for the decorrelation of the dust and synchrotron pattern between $\nu_1$ and $\nu_2$, respectively (see App.~F of \cite{bk15} for details on the decorrelation modeling).\\

The first step of the workflow involves condensing the SAT frequency maps into a set of $B$-mode bandpowers. In its simplest form, this method estimates a total of $N_{\rm freq}(N_{\rm freq}+1)/2 = 45$ auto- and cross-spectra between the $N_{\rm freq}$ bands. 
As discussed in Sec.~\ref{sec:delensing} and \ref{sec:likelihood}, we perform delensing by creating a lensing template and adding it to the CMB-S4 SAT dataset as a pseudo-frequency band, against which cross-spectra are taken \citep{bkspt}. 
For pipeline A, this results in a total of 55 auto and cross-spectra.
To recover the purified $BB$ auto- and cross-power spectra $C_\ell^{BB,\nu_1\nu_2}$ between all frequency maps, we use {\tt S$^2$Hat} \citep{s2hat}, a pseudo-$C_\ell$ estimator that performs $B$-mode purification to account for $E$-to-$B$ leakage \citep{Smith2005,grain09}. 
The breakdown of the auto- and cross-spectra set between different frequencies and across the three foreground models considered in this work is presented in App.~\ref{app:s4_bb_crossfreq} and shown in Fig.~\ref{fig:s4_bb_crossfreq}. 
The bandpowers are extracted in top-hat bins with edges at $\ell_b \in [30, 55, 90, 125, 160, 195]$.
While we do not explicitly include any filtering in our simulations, we limit the analysis to angular scales $\ell \ge 30$ to reflect the modes typically retained after TOD filtering. The recovered auto-spectra are subsequently debiased using an estimate of the average noise power spectrum obtained from noise simulation spectra.
We relate the true modes on the sky to their respective bandpowers in a bin using a bandpower window function computed from simulations. Together with a simulation-based bandpower covariance matrix, we input the computed bandpowers and the model into a Hamimeche-Lewis likelihood \citep{Hamimeche2008}, which is a likelihood approximation of Eq.~\ref{eqn:hl_post}. 

\subsubsection{Pipeline B (Harmonic-space Internal Linear Combination) \label{sec:bILC}}
Pipeline B follows an implementation of the ILC approach, a non-parametric class of component separation algorithms \citep[e.g.][]{Bennett92,Tegmark1996,Bennett_2003,Tegmark2003,eriksen04,Delabrouille_2008,remazeilles10, Leloup2023}.
The ILC method assumes that the CMB signal is statistically independent from non-CMB components.
Under this framework, the data model in Eq.~\ref{eq:data_model} can be understood as a linear mixture of the CMB signal $s_{\rm CMB}$ and a combined noise term $\tilde{\mathbf{n}}$ which aggregates various astrophysical foregrounds and instrumental noise. \\

This pipeline makes use of the estimator in Eq.~\ref{eq:ilcestimator} to construct a foreground-cleaned CMB map, which is the only component to be reconstructed ($N_{\rm comp}=1$). The mixing matrix $\mathbf{F}$ is just a $N_{\rm freq}\times 1$ column vector containing the emission law of the CMB. 
If the input maps are calibrated in thermodynamic units, then its entries are all ones.\footnote{CMB calibration inaccuracies, i.e. $\mathbf{F}\ne \mathbf{1}$, may introduce a multiplicative bias in the recovered CMB signal, see, e.g., \citep{dick10}.} 
The covariance $\mathbf{\hat{C}}$ is estimated using a sample covariance estimator on the input frequency maps $\mathbf{d}$
\begin{equation}
\label{eq:cov_ilc}
\hat{\mathbf{C}}_\ell = \frac{1}{2\ell+1} \sum_m   \mathbf{d}_{\ell m}\mathbf{d}_{\ell m}^\dagger.
\end{equation}

The ILC analysis begins by converting the observed and masked $Q$ and $U$ frequency maps to $E$ and $B$ fields using spin-2 spherical harmonic decomposition.
The purified spherical harmonic $B$-mode coefficients are extracted using the purification scheme \citep{Smith2005,grain09} implemented in the \texttt{NaMaster} software \citep{namaster}.
We then take the spherical harmonic coefficients $a_{\ell m}^{E/B}$ from each channel and convolve them to a common resolution, $a_{\ell m}^{\nu} \to a_{\ell m}^{\nu} b_{\ell}^{\nu_{\rm eff}}/b_{\ell}^{\nu}$ \citep{Rizzieri2024}.
Here, we adopt the common resolution of a Gaussian full-width half-maximum (FWHM) of 22.7 arcminutes. \\

When estimating the covariance matrix $\hat{\mathbf{C}}$ directly from the data, component-separated CMB maps are known to be affected by the so-called ``ILC bias" \citep[see, e.g.,][]{saha08,Delabrouille_2008}. 
This bias arises from chance correlations between the CMB and contaminants such as foregrounds and noise. 
The variance minimization unintentionally cancels out $N_{\rm freq}-1$ CMB modes, leading to a negative bias that is most pronounced on large angular scales.
Various strategies exist in the literature to mitigate the ILC bias, ranging from analytical or simulation-based modeling of the bias \citep[e.g.,][]{saha08} to modifying the cost function minimized in the ILC derivation \citep[e.g.,][]{remazeilles10}. \\

In this work, we follow the approach of \citep{coulton23} to address this bias. The key idea is to ensure that the ILC weights are independent of the data they are applied to, preventing any mode from being weighted by itself.
When estimating the ILC weights for a given mode $m$, we omit that mode from the covariance matrix estimation, as shown in the following expression: 
\begin{equation}
\label{eq:cov_ilc_bias}
\hat{\mathbf{C}}_\ell \to \hat{\mathbf{C}}_{\ell m} \equiv \frac{1}{N_{\rm modes}} \sum_{m' \ne m}  \mathbf{d}_{\ell m'}\mathbf{d}_{\ell m'}^\dagger.
\end{equation}
In practice, when calculating the covariance matrix using Eq.~\ref{eq:cov_ilc_bias}, we average the spherical harmonic coefficients over a bin $b$ of width $\Delta\ell$ centered in $\ell_b$.
Specifically, the covariance matrix is estimated as $\hat{C}_b = \sum_{\ell \in b} \sum_{m' \ne m} \mathbf{d}_{\ell m'}\mathbf{d}_{\ell m'}^\dagger / N_{\rm modes}$, where $N_{\rm modes}$ is the number of $(\ell,m)$ modes within the bin.
For our analysis, we calculate the covariance matrix in linearly spaced bins with a width of $\Delta\ell=100$, starting from $\ell=30$. 
Additionally, we exclude $m$ modes that are separated by $\Delta m = 1$ in the summation. 
Finally, we estimate and subtract the noise bias, $N_\ell^{\rm ILC}$, from the cleaned CMB map by averaging the power spectra of 500 noise-only simulations to which the ILC weights have been applied.\\

Finally, we note two possible extensions of the ILC framework.
First, the harmonic-space ILC algorithm considered here isotropically weights the different frequency maps as
function of scale across the survey footprint. 
While potentially suboptimal, this approach remains effective because the surveyed area is relatively small and situated far from the Galactic plane, which reduces extreme spatial variability in the foreground emission (see Fig.~\ref{fig:mask}).
However, for survey configurations that cover larger portions of the sky, utilizing needlets—a specific family of spherical wavelets—can provide advantageous localization properties in both pixel and harmonic spaces. 
The application of the ILC technique in the needlet frame is referred to as NILC \citep[e.g.,][]{Delabrouille_2008, basak11}.
A second important extension is constrained ILC (cILC), which reconstructs additional components, typically using parametric models, to explicitly null their contribution in the optimal weighting scheme. 
This represents a ``semi-blind" method, allowing the user to ``tune" the level of prior information used \citep[see, e.g.,][]{remazeilles10,abylkairov21}. 
We defer the exploration of these methods to future work.

\subsubsection{Pipeline C (Map-based parametric maximum-likelihood) \label{sec:ML}}

Pipeline C implements a two-step component separation method. 
We first estimate foreground parameters and then reconstruct component maps (Sec.~\ref{sec:flatpriors}). 
In this approach~\citep[see, e.g.,][]{Eriksen2006, Stompor2009}, we first use the likelihood in Eq.~\ref{eq:spectrallikelihood} to solve for the spectral parameters of the foreground model.
Then, we apply the least-squares estimator of Eq.~\ref{eq:mapskysignal} on the frequency maps to solve for the component maps. Similar to pipeline~A we assume $N_\textrm{components}=3$, i.e. the CMB and two Galactic components, dust and synchrotron.
The resulting mixing matrix is that of Eq.~\ref{eq:3componentmixing}.\\

Our implementation of this pipeline works entirely in pixel space with the intention of being easily adaptable to complex inhomogeneous weighting or foreground modeling. In order to simplify the noise covariance we introduce a filtering operator $\mathbf{R}$ that pre-whitens the noisy frequency maps such that $\mathbf{n}$ in our data model has a noise covariance that can be represented by a matrix that is diagonal in pixel space. We achieve this by constructing the filtering operator for each frequency as
\begin{equation}
\mathbf{R}=\mathbf{Y}\frac{1}{\sqrt{1+\left(\frac{\ell}{\ell_{\rm knee}}\right)^\alpha}}\mathbf{Y}^\dagger,
\end{equation}
where $\mathbf{Y}$ and $\mathbf{Y}^\dagger$ are forward and backward spin-2 spherical transformations, and $\ell_{\rm knee}$ and $\alpha$ are parameters taken from the input noise model (see Sec.~\ref{sec:noise}). \\

\section{Simulations \label{sec:skymodel}}

To assess the performance of different component separation algorithms, it is crucial to have accurate simulations of the polarized microwave sky. 
Generating a large number of mock skies is necessary for this task, but computational tractability requires simulations that are reasonably fast to generate. 
Hence, we use a map-based simulation approach, which we describe below.\\

Our simulated skies are constructed by coadding three different components: i) lensed primary CMB; ii) selected Galactic foregrounds and; iii) instrumental noise. 
The signal CMB and foregrounds maps are convolved with Gaussian beams with the  $\theta_{\rm FWHM}$ reported in Tab.~\ref{tab:s4_spec}. 
The maps are generated at a \texttt{HEALPix}\footnote{\url{https://healpix.jpl.nasa.gov}} \citep{healpix} resolution of $N_{\rm side}=512$, corresponding to a pixel size of about 6.9 arcminutes. 
The map based simulations used here are very similar to those used in our previous CMB-S4 forecast paper \citep{s4forecast}.

\subsection{CMB \label{sec:cmb}}
We use CMB signal simulations from the \planck FFP10 suite, generated using the Planck best-fit $\Lambda$CDM parameters \citep{planck_params_2018}: $H_0 = 67.01, {\rm km/s/Mpc}$, $\Omega_c h^2 = 0.1202944$, $\Omega_b h^2 = 0.02216571$, $10^9 A_s = 2.119631$, and $n_s = 0.9636852$. These parameters are fixed throughout the paper.
Unlensed primary CMB skies are simulated as Gaussian realizations of the corresponding power spectra. 
Lensing effects are incorporated by remapping these unlensed realizations using independent Gaussian simulations of the lensing potential $\phi$ with the same input cosmology. 
We generated 500 simulations: 250 with a tensor-to-scalar ratio of $r = 0.003$ and 250 with $r = 0$, allowing us to assess the impact of a minimal primordial tensor signal.

\subsection{Foregrounds \label{sec:fg}}
Two main astrophysical foregrounds are relevant for CMB polarization studies at large and intermediate angular scales: Galactic dust and synchrotron.
Interstellar dust grains in the Milky Way absorb starlight and re-radiate it as thermal emission in the far-IR band, dominating the foreground emission at frequencies $\nu \gtrsim 60$~GHz.
Synchrotron emission is due to relativistic cosmic-ray electrons spiraling around the magnetic fields in our galaxy, representing the dominant diffuse component at $\nu \lesssim 60$~GHz.
To validate and evaluate the performance of our component separation algorithms, we generate three distinct suites of foreground simulations, each with progressively increasing complexity—ranging from purely Gaussian, statistically isotropic foreground emissions to complex models that capture the non-Gaussian structure of the Galactic magnetic field.
We then apply the pipelines on simulations with \texttt{PySM} foreground models.

\subsubsection{Model 0 (Gaussian foregrounds)}
Model 0 (Gaussian foregrounds) is a suite of simple isotropic Gaussian simulations that include emission from Galactic dust and synchrotron. 
We simulate foreground maps at 353 GHz for dust and 23 GHz for synchrotron using power-law input $BB$ power spectra, given by $\mathcal{D}^{BB, f}_{\ell}=A_f \left(\frac{\ell}{80}\right)^{\alpha_f}$, where $f\in \{d,s\}$ indicates the foreground type, $A_f$ represents the amplitude parameter at the pivot scale of $\ell = 80$, and $\alpha_f$ is the spatial spectral index. 
The fiducial parameter values used to generate these foreground simulations are set to $A_s = 3.8\,\mu \text{K}^2$ and $A_{\rm d} = 4.25\,\mu $K$^2$ for the amplitude of synchrotron and dust, respectively. 
Correspondingly, the spatial spectral parameters are \( \alpha_{\rm d} = -0.4 \) for dust and \( \alpha_{\rm s} = -0.6 \) for synchrotron. The SED parameters are fixed to \( \beta_{\rm d} = 1.6 \) for dust and \( \beta_{\rm s} = -3.1 \).
These parameter values are similar to or consistent with those found in the BICEP/Keck field \citep{BKVI}.\\

To simulate the foreground maps at different frequencies, we scale the dust and synchrotron maps using their respective SEDs of Eqs.~\ref{eq:dustSED} and \ref{eq:syncSED}, which are assumed to be uniform over the sky. These simulations include the (Gaussian) sample variance since each foreground map is generated with a different seed.

\subsubsection{Model 1 (Amplitude modulated Gaussian foregrounds)}
This foreground model is an extension of model 0 introducing spatial modulation of the foreground brightness. To achieve this, we multiply the Gaussian realizations from model 0 by a template that captures the spatial variations in foreground amplitudes. 
By introducing spatial modulation of the dust and synchrotron amplitudes, this model can better represent the complexity and realism of the foreground emissions. 
We construct this template empirically from the \planck 353 GHz data using power-law fits to the $BB$ power spectra of circular patches of approximately 400 deg$^2$ centered on each of the $N_{\rm side}=8$ pixel centers of an \texttt{HEALPix} map.
The resulting map is normalized by the fiducial dust amplitude and smoothed with a 10-degree Gaussian taper. This position-dependent rescaling of the foreground maps introduces some non-Gaussianity. Each dust and synchrotron map from model 0 is scaled by the same data-driven template. 
As in model 0, the SED is assumed to be uniform across the sky.

\subsubsection{Model 2 (Vansyngel model)}
The final model we consider for pipeline validation is based on the work of \citet{Vansyngel2016}. 
This phenomenological model provides a framework for  polarized emission of dust and synchrotron in the sub-millimeter band using the structure of the Galactic magnetic field and interstellar matter.
The model assumes that the Galactic magnetic field is the result of a superposition of a mean uniform field and a three-dimensional Gaussian random turbulent component with a power-law spectrum. 
The gas density distribution in the diffuse interstellar medium is derived from the total intensity of dust as observed by \planck. 
These simulations are tuned to match the foreground properties observed by \planck and incorporate non-Gaussian features as well as spatial variations in  $\beta_{\rm d}$, which directly cause frequency decorrelation \citep[e.g.,][]{planck_int_L}.
By incorporating the structure of the Galactic magnetic field and the interstellar matter, this model attempts to capture more realism of the foreground emissions.
Note that for this specific foreground model there is only one realization such that every map in our simulation suite contains the exact same foreground component.

\subsubsection{\texttt{PySM} models}\label{sec:pysm_models}
To apply and test the foreground-cleaning methods as if working with real data in Sec.~\ref{sec:results_pysm}, we also consider three separate microwave sky simulations using foreground templates of varying complexities provided by the \texttt{PySM3}\footnote{\url{https://pysm3.readthedocs.io}} package \citep{pysm3} (as opposed to our previous work \citep{s4forecast} which used \texttt{PySM2}). These templates have different properties compared to the three models we use to validate the pipelines.
The models are labeled as follows, with \texttt{PySM}-internal nomenclature in brackets:
\begin{itemize}
\item Model 3 (\texttt{d9, s4}): Large-scale Galactic thermal dust SED is modelled as a single-component modified blackbody with fixed spectral index of $\beta_{\rm d} = 1.48$ and temperature of $T_{\rm d}=19.6$ K, with templates derived from \planck GNILC maps \cite{planck18_diffuse_compsep} (dust model \texttt{d9} in \texttt{PySM}). 
Similarly, the Galactic synchrotron SED is assumed to be uniform over the sky with a constant spectral index of $\beta_{\rm s}=-3.1$ (\texttt{s4}) and a spatial template from WMAP 9 year 23 GHz polarization maps. 

\item Model 4: (\texttt{d10, s5, co3}): This intermediate complexity model integrates \texttt{d10} for dust with small-scale fluctuations and high-resolution templates derived from \planck GNILC maps with spatial variations in $\beta_{\rm d}$ and $T_{\rm d}$. 
The synchrotron templates are the same as \texttt{s4} but the spectral index map is based on the S-PASS data \cite[\texttt{s5},][]{Krachmalnicoff18}. 
Non-Gaussian small-scale fluctuations are introduced in the template maps using the polarization fraction tensor formalism, while Gaussian small-scale fluctuations are added to the spectral parameter maps as realizations of power-law power spectra.
We include CO polarized emission at the level of 0.1\% and simulated CO clouds \citep[\texttt{co3},][]{puglisi17}.

\item Model 5 (\texttt{d12, s7, a2, co3}): The most complex suite includes the MKD 3D model of polarized dust emission with 6 layers, each with different templates, spectral index and dust temperature \cite[\texttt{d12},][]{MKD}.
For synchrotron, \texttt{s7} builds on \texttt{s5} and adds a curvature term to the SED based on the ARCADE experiment results \citep{Kogut12}.
Free-free and CO emission as in model 4 and  \texttt{a2} assumes a sky-constant 2\% polarization fraction for the polarized anomalous microwave emission (AME).
\end{itemize}

While these models reflect increasing complexities in their modeling, and are anchored on Planck data in different ways, the properties of the components are not always consistent with those measured by experiments on subpatches~\citep[see e.g.][]{spider2024}. 
However, for our purposes of pipeline testing, including a range of models with lower to higher levels of complexity is informative to our understanding of the limits of each pipeline.

\subsection{Instrumental noise \label{sec:noise}}
Map-level noise simulations for the SATs are synthesized following the scheme outlined in \citep{s4forecast}.
This involves converting BICEP/Keck (BK) noise bandpowers into a prescription for map noise.
To generate the maps, we first fit the noise curves $N_{\ell}$s at each frequency from BK to a $1/f$ noise model
\begin{equation}
\label{eq:cmb_1of_noise}
    N_{\ell}=\Delta_P^2\left[1+\left(\frac{\ell}{\ell_{\rm knee}}\right)^{\alpha_{\rm knee}}\right],
\end{equation}
accounting for observational effects such as beam smoothing and timestream filtering. 

As shown in Tab.~\ref{tab:s4_spec}, the resulting knee multipoles for the South Pole SATs are  $\ell_{\rm knee} = 60$ (except for the 20 GHz channel that has $\ell_{\rm knee}=150$ since it is observed by the LAT) while the slope of the power-law noise component varies from $-1.7$ (30 to 95~GHz) to $-3.0$ (145 to 270~GHz).
In Tab.~\ref{tab:s4_spec}, we report the white noise levels across the used bands.
Then, the noise levels are rescaled by the ideal ratios between the Noise Equivalent Temperature (NET) from BK and CMB-S4, as well as the number of detectors-years ratios,\footnote{Note that for the 20 GHz channel, which is placed on the delensing LAT, we use different shape parameters $(\ell_{\rm knee},\alpha_{\rm knee})$ based on SPT measurements. }
\begin{equation}
\sigma_{\text {map,S4 }}=\sigma_{\text {map,BK }} \frac{\mathrm{NET}_{\mathrm{S} 4, \text { ideal }}}{\mathrm{NET}_{\mathrm{BK}, \text { ideal }}} \sqrt{\frac{N_{\text {det-yr,BK }}}{N_{\text {det-yr,S4 }}}},
\end{equation}
where the map depth $\sigma_{\rm map}$ is obtained from the white noise levels as $\sigma_{\rm map} = \frac{180\times60}{\pi}\sqrt{\Delta_P}$.
Using these scaled noise curves $N_{\ell}$s, we generate Gaussian noise realizations.
Finally, the noise maps are divided by the square-root of the coverage map (i.e. the hit map, see Sec.~\ref{sec:mask}) to boost the noise around the edges of the survey footprint, similarly to what would be observed in real data.
We note that the simulations do not model the effects of time-stream filtering.

\begin{table}[t]
\centering
\begin{tabular}{c|c|c|c|c}
\toprule
\rowcolor{gray!30} Frequency & $\theta_{\mathrm{FWHM}}$ & Noise $\Delta_P$ & $\ell_{\mathrm{knee}}$ & $\alpha_{\mathrm{knee}}$ \\
\rowcolor{gray!30} (GHz) & (arcmin) & ($\mu$K-arcmin) & & \\
\midrule
20  & 11  & 13.6  & 150 & -2.7 \\
30  & 73  & 3.53  & 60  & -1.7 \\
40  & 73  & 4.46  & 60  & -1.7 \\
85  & 26  & 0.88  & 60  & -1.7 \\
95  & 23  & 0.78  & 60  & -1.7 \\
145 & 26 & 1.23  & 60  & -3.0 \\
155 & 23 & 1.34  & 60  & -3.0 \\
220 & 13 & 3.48  & 60  & -3.0 \\
270 & 13 & 5.97 & 60  & -3.0 \\
\bottomrule
\end{tabular}
\caption{Technical specifications for a CMB-S4 Small Aperture Telescope (SAT). We show the frequency band centers, FWHM apertures, polarization white-noise levels, and atmospheric noise parameters. The total noise power spectrum is modelled as $N_{\ell}=\Delta_P^2\left[1+\left(\ell/\ell_{\mathrm{knee}}\right)^{\alpha_{\mathrm{knee}}}\right]$.The noise levels reflect refinements to the CMB-S4 Conceptual Design Report (CDR) specifications, preceding updates introduced in the Preliminary Baseline Design Review (PBDR).}
\label{tab:s4_spec}
\end{table}

\subsection{Mask \label{sec:mask}}
\begin{figure}[t]
    \centering
    \includegraphics[width=\columnwidth]{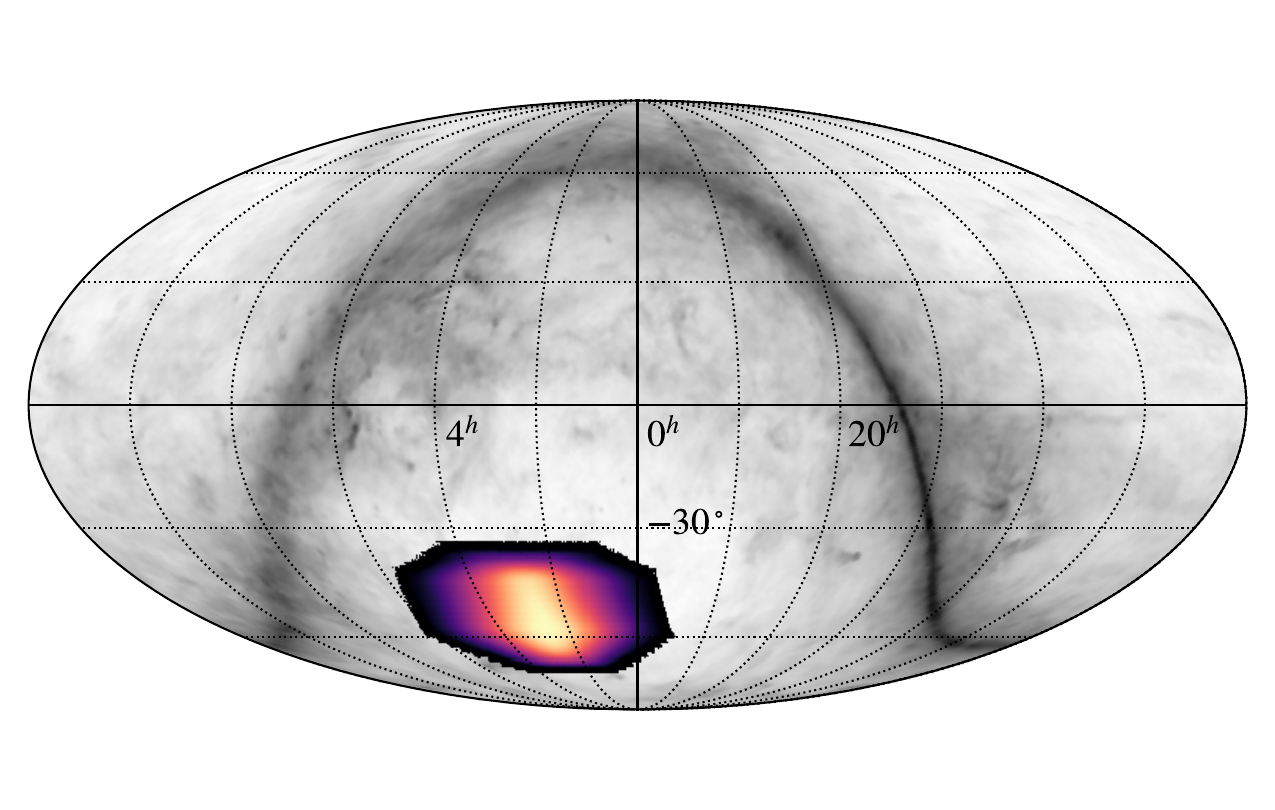}
    \caption{Normalized hit pattern on the sky for the South Pole Small Aperture Telescopes' ultra-deep survey. The grey background shows the Galactic dust map from \textit{Planck} \texttt{Commander} in intensity plotted on a logarithmic scale.}
    \label{fig:mask}
\end{figure}

The SAT ultra-deep survey, as envisioned from the South Pole, uses constant elevation scans to observe a patch that is 3\% of the sky.
The mask considered in this work is constructed from a smoothed version of the hits count map corresponding to this scanning strategy and is shown in Fig.~\ref{fig:mask}.

\section{Delensing}
\label{sec:delensing}
Below instrumental noise levels of about 5 $\mu$K-arcmin, lensing $B$ modes becomes a more dominant source of sample variance, thus limiting the improvement of the constraining power on $r$ \citep[e.g.,][]{Knox_2002,Kesden_2002,Seljak_2004}.
Delensing techniques have been developed to remedy this issue  \citep[e.g.,][]{Smith_2012,Sherwin_2015,Carron2017,Millea_2019}.
Recent work has highlighted the suitability of delensing techniques for combining data from various cosmological probes, such as ground- and space-based CMB experiments \citep{Lonappan2024} and large-scale structure surveys \citep{Namikawa2024}.
The basic idea underlying delensing is to use low-noise observations of $E$ modes and estimates of the lensing potential $\phi$ over the same patch of the sky to reconstruct the specific lensing-induced $B$ modes in that region.
In this work, we implement delensing by using the curved-sky extension of the iterative algorithm presented in Ref.~\citep[][]{Carron2017}.
A detailed description and validation of the delensing algorithm is presented in Ref.~\citep{belkner2023cmbs4} while here we provide a brief summary of its key aspects. \\

We use the low-noise, high-resolution simulated maps from the South Pole LAT, which share the same signal and foregrounds as the SATs but different noise realizations and beam sizes, to predict the lensing-induced CMB polarization, also referred to as the lensing $B$-mode template (LT), $B^{\rm LT}$. 
Schematically, this template can be written as a remapping of the unlensed $E$-mode-only map operated by the deflection vector field $\boldsymbol{\alpha}$ (i.e. the gradient of the CMB lensing potential, $\boldsymbol{\alpha}=\nabla \phi$):
\begin{equation}
\hat{B}^{\rm LT} \equiv \hat{\boldsymbol{\alpha}}^{\rm MAP} \circ \hat{E}^{\rm unl}.
\end{equation}
Here, $\boldsymbol{\alpha}^{\rm MAP}$ is the MAP deflection field (or ``the most probable CMB lensing map") while $\hat{E}^{\rm unl}$ is the Wiener-filtered $E$-mode map which represents our best estimate of the unlensed $E$ modes.\\

The first step of the pipeline combines the LAT maps at different frequencies (with the exception of the 20~GHz channel) using an ILC approach to produce a foreground-cleaned CMB map.
The ILC weights $w_{\ell}^{\nu}$ are obtained using a cross-frequency data covariance matrix modeled with the analytical input foreground power spectra from our Gaussian model 0 (see Sec.~\ref{sec:bILC}).
The foreground-cleaned CMB maps are fed to the delensing pipeline, which iteratively maximizes the lensing map posterior, $\log{P}(\boldsymbol{\alpha}|X^{\rm dat})$, where $X^{\rm dat}$ are the $Q/U$ polarization maps. 
At each step, the gradients of the CMB lensing likelihood and prior are calculated and used to progress towards the MAP lensing map, using a variant of the Newton–Raphson optimization method.
The resulting delensing efficiency reaches approximately 92\% depending on the specific foreground model.\\

In this analysis we reconstruct the LT using polarization data alone for three main reasons: 
i) the expected gain from adding temperature data is minimal\footnote{For the South Pole deep patch considered in this work, \citep{belkner2023cmbs4} find that including temperature maps in the reconstruction reduces the the residual lensing power only by about 0.2\% (assuming that the impact of foregrounds and atmospheric noise is under control).} at these low noise levels;
ii) including temperature data leads to a stronger ``mean-field" term; and 
iii) temperature data are substantially affected by extragalactic foreground emission (especially the Sunyaev-Zel'dovich effects and cosmic infrared background) at high-$\ell$, which significantly complicates the analysis.
In addition, we remove any $B$-mode power below $\ell_B <200$ from the LAT maps prior to the LT reconstruction to avoid any potential spurious internal delensing bias caused by the overlap with the modes probed by the SATs \citep[e.g.,][]{Teng11,Beck2020,Han2020,BaleatoLizancos2020}.\footnote{Specifically, we restrict the multipole range to $\ell_E \in [2,4000]$ and $\ell_B \in [200,4000]$ for the $E$- and $B$-modes, respectively.}\\

We incorporate the reconstructed LT into the cosmological inference framework in a similar fashion to the joint analysis of BICEP/\textit{Keck} and SPT data presented in \citet{bkspt}.
Specifically, for pipeline A, the extracted LT is added as a pseudo-frequency band to the CMB-S4 SATs dataset, against which cross-spectra are taken. 
For pipelines B and C, the LT and the cleaned CMB map are the input maps used to reconstruct the spectra that are fed to the likelihood.

We need a statistical characterization of the signal and noise components of the LT auto-spectrum as well as the cross-spectra with both the SAT frequency and cleaned CMB maps. 
Therefore, we model the LT as a noisy filtered version of the true lensing $B$ mode plus an additive noise term:
\begin{equation}
\hat{B}_{\ell m}^{\rm LT} = \alpha_{\ell} (B_{\ell m}^{\rm lens} + n_{\ell m}^{\rm LT}), 
\end{equation}
where $\alpha_{\ell}$ is an effective isotropic filter function which we estimate by correlating 500 signal-only lensed-$\Lambda$CDM (LLCDM) input simulations with the LT,
\begin{equation}
\alpha_\ell = \frac{\hat{C}_\ell^{\rm LLCDM \times LT}}{\hat{C}_\ell^{\rm LLCDM \times LLCDM}}.
\label{eqn:alphaell}
\end{equation}
The corresponding effective LT noise auto-spectrum can then be calculated as:
\begin{equation}
N_\ell^{\rm LT} = \frac{\hat{C}_\ell^{\rm LT \times LT}}{\alpha_\ell^2} - \hat{C}_\ell^{\rm LLCDM \times LLCDM}.
\label{eqn:ltnoise}
\end{equation}
In App.~\ref{app:lt_validation}, we provide a validation of the LT.
Specifically, in Fig.~\ref{fig:lt_plots} we show the LT noise power spectra measured from the three foreground suites and the average cross-correlation between the cleaned CMB maps recovered by pipeline B and C and the lensing template.\\

Given Eq.~\ref{eqn:ltnoise}, the LT noise term includes any effect that is present in the simulations used to calculate the LT auto-spectrum $\hat{C}_\ell^{\rm LT \times LT}$. 
In our current analysis, we use the same full-signal simulation inputs for both the LT and SAT maps, meaning they include CMB, instrumental noise, and foregrounds. 
Notably, for the Vansyngel and \texttt{PySM} models (2-5), the foregrounds are fixed across all realizations. 
Therefore, foregrounds leaked into the LT will be exactly captured by $N_\ell^{\rm LT}$.
In real data, however, it will not be possible to characterize the $N_\ell^{\rm LT}$ term as precisely.
In App.~\ref{app:lt_validation}, we show that the bias in $N_\ell^{\rm LT}$ from higher-order correlations of foregrounds is negligible by comparing the difference in $N_\ell^{\rm LT}$ between the fiducial set of simulations and a matching set with only Gaussian foregrounds. 
Given the negligible impact on $r$, we proceed to use the fiducial foreground simulations in constructing $N_\ell^{\rm LT}$ as opposed to using an appoach closer to what can be applied to data presented in App.~\ref{app:lt_validation}.


\section{Bandpower likelihoods and component/residual models}
\label{sec:likelihood}
%
A bandpower-level likelihood is used by all three pipelines to constrain the tensor-to-scalar ratio $r$. 
These likelihoods utilize bandpowers extracted from a given realization, along with model, fiducial, and noise bandpowers, and the bandpower covariance matrix. Pure-$B$ bandpowers are extracted using either {\tt NaMaster} (pipelines B and C) or {\tt S$^2$Hat} (pipeline A) \citep{namaster,s2hat}.\\

The CMB theory model is common to all three pipelines. 
It includes the lensing $B$ modes $C^{BB,\rm lens}_{\ell}$ from the $\Lambda$CDM model, described by the parameters that generate the CMB simulations (Sec.~\ref{sec:cmb}), and the tensor modes $C^{BB,\rm ten}_{\ell}$ for different $r$ values.\\

For Pipeline A, $r$ is jointly estimated with all foreground parameters by using the Hamimeche-Lewis likelihood ~\cite{Hamimeche2008}, an approximation of the posterior given in Eq.~\ref{eqn:hl_post}. 
The inputs to this likelihood are frequency auto- and cross-spectra. The theory model of the CMB, dust, and synchrotron components in the frequency map bandpowers is detailed in Sec.~\ref{sec:crossCl}. 
The theory $C_{\ell}$s are generated using \texttt{CAMB}~\cite{camb} and binned by the bandpower window function, which is computed by running simulated maps with non-zero power in only a single multipole $\ell$ at a time, to be consistent with the analysis in Ref.~\cite{s4forecast}. 
The fiducial model bandpowers are constructed from input CMB theory, dust, and synchrotron spectra.
The noise bandpowers per-frequency are estimated from simulation means and the bandpower covariance matrix is constructed semi-analytically and conditioned as detailed in Appendix B of~\citet{s4forecast}. \\

For Pipelines B and C the inputs are auto- and cross-spectra of the CMB component and the lensing template. 
The model in these likelihoods only includes a CMB component and potential foreground residuals in the CMB component map. 
The theory model of the lensing template bandpowers and the CMB $\times$ LT bandpowers are given by $\alpha_{\ell}^2 C^{BB, \rm lens}_{\ell}$ and $\alpha_{\ell} C^{BB,\rm lens}_{\ell}$, respectively, where $\alpha_{\ell}$ is the isotropic filter function defined in Eq.~\ref{eqn:alphaell}. 
The theory $C_{\ell}$s are generated using \texttt{CAMB}~\cite{camb} and binned by the bandpower window function output from {\tt NaMaster}, given the mask applied to the maps.
The fiducial model bandpowers are taken to be the simulation mean from the model 0 simulation set with $r=0$.
The noise bandpowers are estimated from the mean of each set of simulations for the CMB component, given by Eq.~\ref{eqn:ltnoise} for the LT auto-spectra, and set to zero for the CMB $\times$ LT cross-spectra. 
The bandpower covariance matrix is estimated from simulation bandpowers from the model 0 simulation set.
We apply the Hartlap-Anderson factor~\cite{hartlap07} to remove the parameter bias caused by covariance inverses estimated from a finite number of simulation realizations.
We condition the covariance matrix by zeroing elements beyond the closest neighbor to the diagonal to avoid misestimation of the marginalized $r$ posterior \cite{Beck2022}. 
We use \texttt{Cobaya}~\cite{cobaya} to search for the maximum a posteriori $r$. \\

For the Model 2 simulation suite, the CMB components extracted from both Pipelines B and C are contaminated by foreground residuals. 
We model the foreground residuals as part of the model bandpowers in the likelihood.
For Pipeline B, the foreground residuals are modeled by applying the corresponding ILC weights to the foreground-model power spectra of Pipeline A (Eq.~\ref{eqn:bk_fgmodel}),
\beq
\label{eq:fg_marg_ilc}
\mathcal{D}_{\ell}^{BB, {\rm fgres}} = \sum_{\nu_1 \nu_2} w_{\nu_1} w_{\nu_2} \mathcal{D}_{\ell}^{BB,\nu_1 \nu_2},
\eeq
where $w_{\nu_1}$ and $w_{\nu_2}$ are derived per realization and, unlike the ILC weights applied to the maps, no $m$-mode masking is needed. 
We vary $A_{\rm d}$ and $\alpha_{\rm d}$ while fixing the rest of the foreground parameters to the best fits to each realization from Pipeline A. 
For Pipeline C, the foreground residuals are modeled to be a linear combination of the power spectra of the respective dust and synchrotron components for each realization,
\beq
\label{eq:fg_marg_ml}
C_{\ell}^{BB,{\rm fgres}} = A_{\rm d}' C_{\ell}^{BB, {\rm dust}} + A_{\rm s}' C_{\ell}^{BB, {\rm sync}},
\eeq
where $A_{\rm d}'$ and $A_{\rm s}'$ are also varied in the likelihood. \\

\section{Results and discussion} \label{sec:results}
This section presents and discusses the main results of our foreground cleaning method comparison. 
We begin with a visual comparison of the cleaned CMB $B$-mode maps from Pipelines B and C (Sec.~\ref{sec:clean_cmb_maps}), followed by an assessment of the cleaned CMB spectra and cosmological inference, including biases and uncertainties on $r$, for different cleaning methods and foreground simulations (Sec.~\ref{sec:clean_cmb_spectra} and \ref{sec:r_constraints}). 
We conclude with a discussion of results based on the \texttt{PySM} foreground suites (Sec.~\ref{sec:results_pysm})

\begin{figure*}[t]
    \centering
    \includegraphics[width=\textwidth]{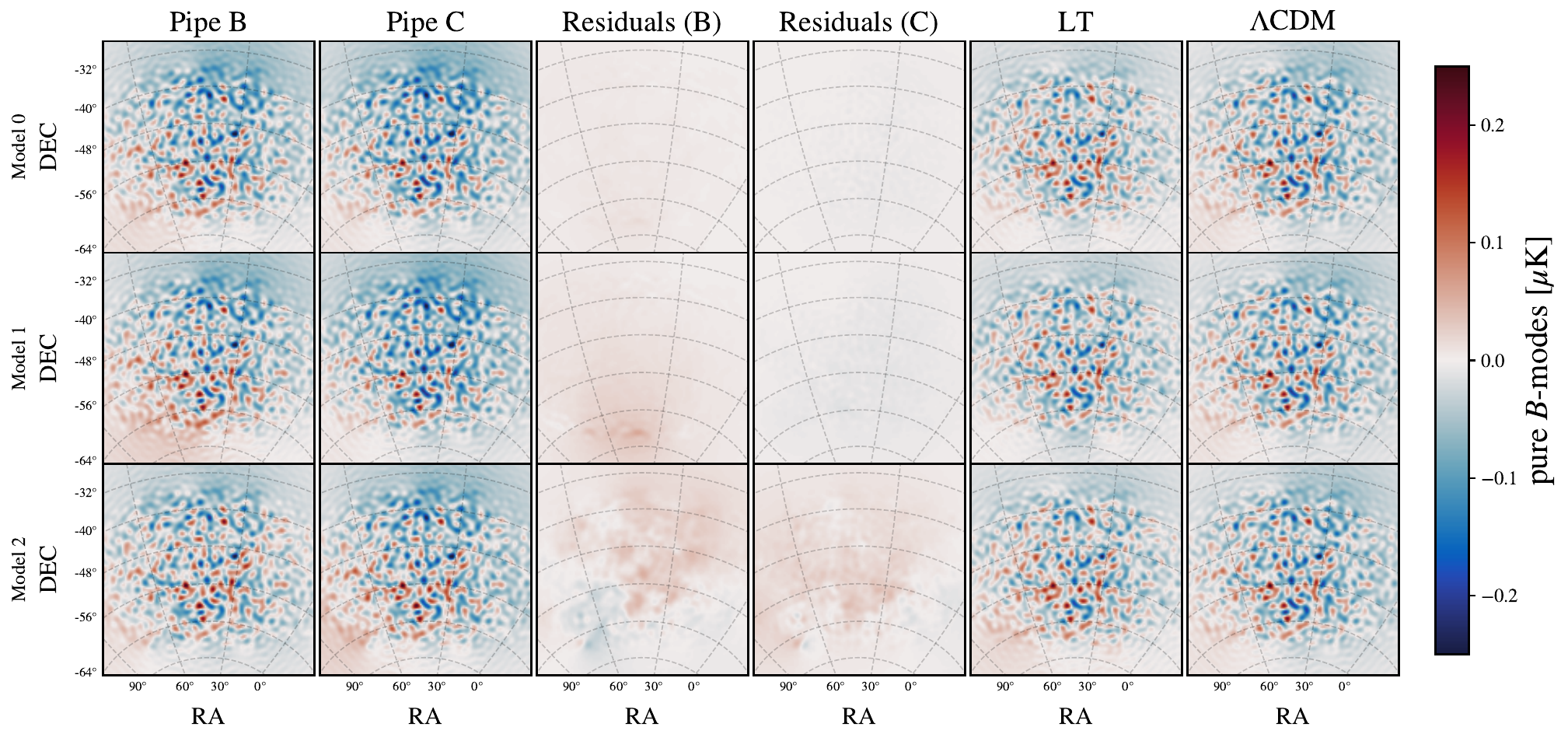}
    \caption{Comparison of $B$-mode maps and related map products from Pipelines B and C. The first two columns present the cleaned CMB maps from Pipelines B and C, respectively. The central panels display the foreground residuals, calculated as the difference between the cleaned total CMB map and the combined lensing $B$ modes and noise maps after component separation. The two columns on the right illustrate the lensing template and the input lensed CMB maps. Different rows correspond to different foreground models, ranging from Model 0 (top) to Model 2 (bottom). All maps are band-limited to $30 \le \ell \le 200$ to highlight scales relevant for $r$ inference.
    With Model 0 (Gaussian foregrounds), the residuals are negligible from both pipelines, as expected. With Model 2, the residuals show spatial variations and differences between methods---reflecting the differences in the assumptions embeded in each foreground-cleaning method. The LT and $\Lambda$CDM panels provide a visual confirmation of negligible foreground contaminations in the LTs. }
    \label{fig:maps}
\end{figure*}

\subsection{Cleaned CMB maps \label{sec:clean_cmb_maps}}
First, we provide a visual comparison of the methods that produce a cleaned reconstruction of the CMB signal, namely pipelines B and C.
Visualizing the overall amplitude and spatial dependence of the map-level residuals across different configurations can be useful to build intuition regarding the response of the component-separation methods (and their assumptions) to the complexity of foregrounds.\\

In Fig.~\ref{fig:maps}, we show $B$-mode maps of a representative realization drawn from our simulation suites.
Each one of the three rows corresponds to a given foreground model, from Model 0 to Model 2 (top to bottom).
Along columns we show the input cosmological signal (which is shared across the simulations suites), the cleaned CMB maps recovered by Pipelines B and C, their respective residuals computed against the input $B$ modes, and the lensing template reconstructed from the corresponding LAT data.
We observe the foreground residuals are substantially smaller than the amplitude of the input lensing $B$-mode fluctuations, suggesting that the component-separation methods are indeed cleaning the observed maps.
The overall amplitude of residuals increases with the complexity of the foreground models, with evidence indicating that the contamination primarily arises from large-scale modes.

\subsection{Power spectra \label{sec:clean_cmb_spectra}}
\begin{figure*}[]
    \centering
    \includegraphics[width=\textwidth]{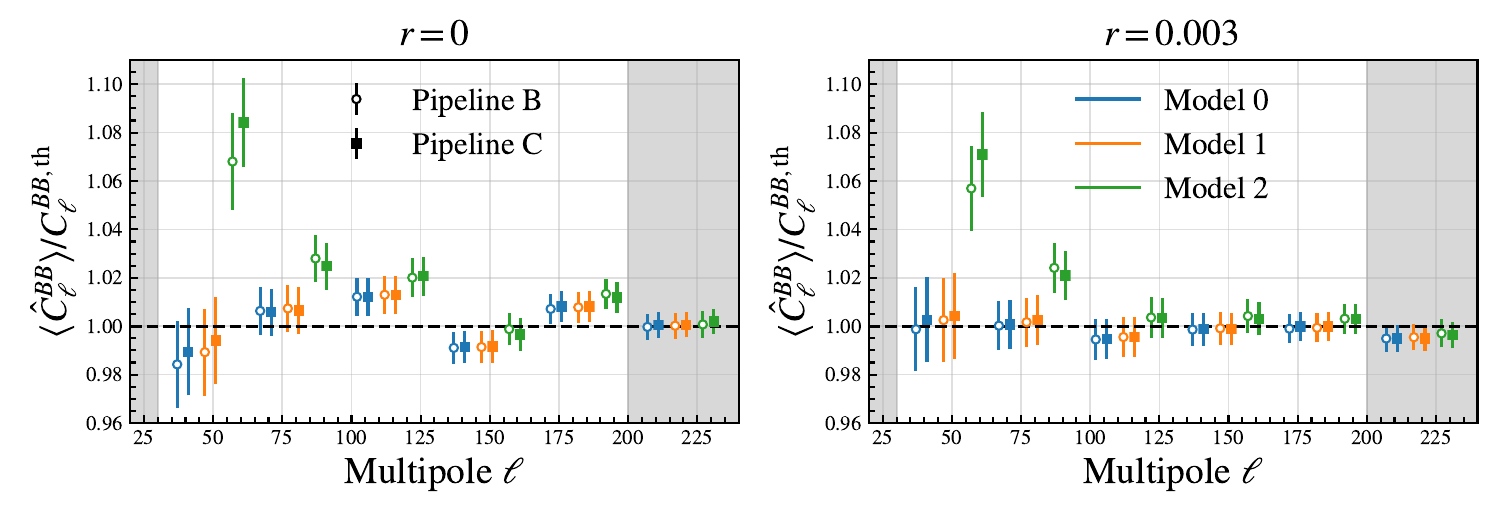}
    \caption{Comparison of the recovered mean cleaned CMB $BB$ power spectrum across the two component-separation methods (B and C) and across foreground suites (0/1/2), normalized with respect to the input theory. In each panel, the empty circle points show the results from Pipeline B, while the filled squares refer to Pipeline C. We show results obtained for foreground suites 0, 1, and 2 in blue, orange, and green, respectively. The left panel shows results for $\Lambda$CDM-only simulations ($r=0$), while the right panel presents the case with PGW power ($r=0.003$). Note that the error bars are divided by $\sqrt{250}$. The shaded gray bands denote modes at $\ell < 30$ and $\ell > 200$, which are not used in the likelihood analysis.
    As in Fig.~\ref{fig:maps}, Model 2 shows a notable excess of residual power at large angular scales, whereas Models 0 and 1 exhibit minimal to negligible residuals.}
    \label{fig:spectra_comp_ml_vs_ilc}
\end{figure*}
We next analyze the CMB power spectra recovered by pipelines B and C. Figure~\ref{fig:spectra_comp_ml_vs_ilc} compares the mean CMB-only power spectra for different foreground models and these two pipelines.\footnote{Pipeline A could, in principle, be modified to recover CMB-only bandpowers from multi-frequency power spectra by marginalizing over foregrounds with an MCMC sampler, as demonstrated in, e.g., Ref.~\citep{wolz24}.}
The figure is divided into two panels: the left panel corresponds to a tensor-to-scalar ratio $r=0$; and the right panel corresponds to $r=0.003$. 
In each panel, we display the instrumental noise-debiased $BB$ power spectrum derived from the cleaned CMB maps. 
These results are averaged over 250 simulated skies, and the error bars are divided by $\sqrt{250}$ to reflect the scatter around the mean. 
We normalize all mean spectra against the corresponding binned input theoretical models. 
Specifically, the left panel is normalized to a lensed-$\Lambda$CDM power spectrum with no PGW contribution ($r=0$), while the right panel is normalized to a lensed-$\Lambda$CDM model that includes tensor power ($r=0.003$).
To facilitate comparison, we use distinct colors to represent the various foreground simulation suites, and markers to differentiate between the component separation methods.\\

The results show that for the Gaussian (Model 0) and amplitude-modulated (Model 1) foreground models, the recovered mean CMB spectra from both pipelines B and C do not exhibit significant bias.
However, for the more complex foreground suite (Model 2), we observe an excess of power at larger angular scales ($\ell \lesssim 100$). 
This excess in Model 2 can be attributed to the spatially varying SED of the foreground components. 
In the ILC approach (Pipeline B), the weights used for component separation are averaged spatially, limiting the method’s ability to capture local variations. 
Similarly, the map-based parametric maximum-likelihood pipeline C is limited in this paper to a fixed dust spectral index $\beta_{\rm d}$ across the entire patch, which further reduces flexibility. 
As a result, neither pipeline is sufficiently flexible to describe the complexity of foregrounds in this suite, leading to the observed excess in the power spectra.
In all cases, the differences in bandpowers between pipelines B and C across the various foreground suites remain well within $1\sigma$, indicating similar performance.

\subsection{Bias and uncertainty of the tensor-to-scalar ratio $r$ estimation \label{sec:r_constraints}}
\begin{figure}
    \includegraphics[width=\columnwidth]{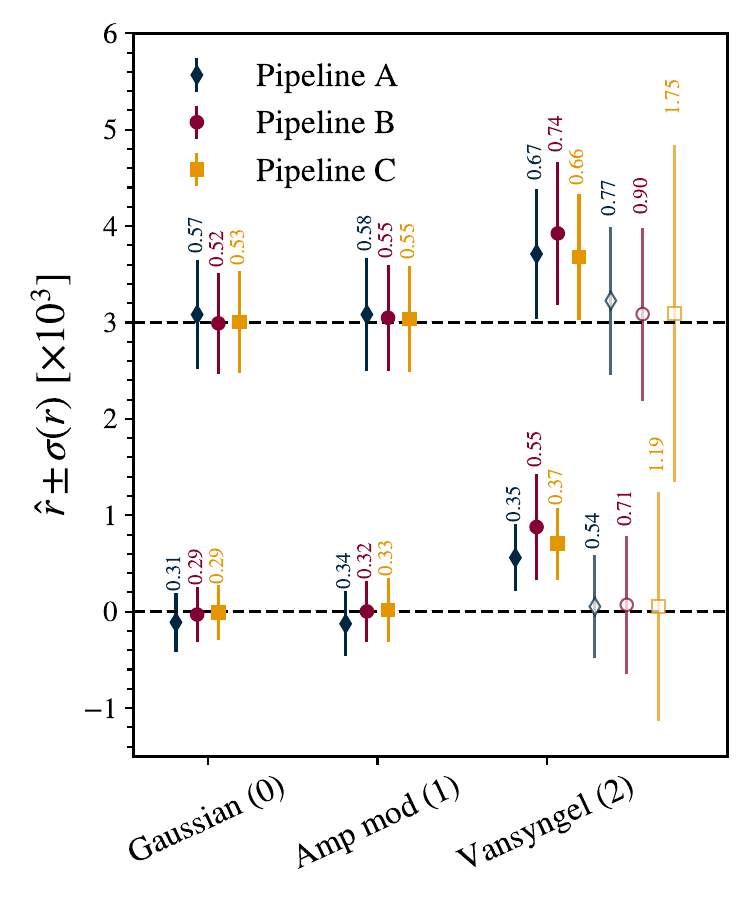}
    \caption{Summary of the biases and uncertainties on the inferred tensor-to-scalar ratios across component-separation pipelines and foreground models. Each point shows the mean $\bar{r}$ and standard deviation $\sigma(r)$ (in units of $10^{-3}$). Results are shown for foreground models 0, 1, and 2 from left to right, while the blue diamond/red circle/orange square markers indicate pipelines A, B, and C, respectively. Empty markers show results from the extended cleaning pipelines (only for the high-complexity Model 2). Points at the bottom correspond to input $r = 0$, while those at the top are for $r = 0.003$.}
    \label{fig:r_bias_unc}
\end{figure}

Given the three pipelines introduced above we can estimate the best-fit $r$ for each simulation realization for each pipeline. This allows us to estimate the bias ($\bar{r}$) and sensitivity ($\sigma(r)$) for each pipeline variation and foreground model by calculating the simulation ensemble mean and standard deviation, respectively; they are summarized in Fig.~\ref{fig:r_bias_unc} and Tab.~\ref{tab:r_bias_unc}. 
These results take into account the delensing processes described in Sec.~\ref{sec:delensing}.\\

We first focus on the $r=0$ scenario.
All three component-separation methods when applied to the Gaussian foreground suite (model 0) yield unbiased estimates of $r$ with comparable statistical uncertainties, $\sigma(r)\simeq 3 \times 10^{-4}$. This is expected and serves as a pipeline validation test.
Pipelines B and C return a mean $r$ consistent with the input value that is within $0.1$ of the statistical uncertainties $\sigma(r)$, while the one from pipeline A is consistent at the $0.35\sigma$ level.
Foreground non-Gaussianity caused by a spatially varying dust amplitude as in Model 1 leads to unbiased $r$ estimates in all pipelines. 
The higher overall foreground levels in Model 1 increase the variance in $r$ by about 13\%. The more complex non-Gaussian foreground Model 2, however, introduces biases between $1.6\sigma$ and $1.9\sigma$ across the methods, with only minor increases in statistical uncertainties for Pipelines A and C, but a notable $\sigma(r)$ increase of about 70\% for Pipeline B. \\

\begin{table}[t]
\renewcommand{\arraystretch}{1.5} 
\begin{tabular}{ccccc}
\toprule
\rowcolor{gray!30} \multicolumn{5}{c}{$\bar{r} \pm \sigma(r) \, [\times 10^3]$}                                                                                                                                               \\ \hline
\hline
\multicolumn{1}{c|}{\multirow{2}{*}{Input $r$}} & \multicolumn{1}{c|}{\multirow{2}{*}{FG model}}     & \multicolumn{1}{c|}{Pipeline A}   & \multicolumn{1}{c|}{Pipeline B}               & Pipeline C                \\
 \multicolumn{1}{c|}{}                           & \multicolumn{1}{c|}{}                              & \multicolumn{1}{c|}{(ext pipe)}      & \multicolumn{1}{c|}{(ext pipe)}      & (ext pipe)      \\ \hline
\multicolumn{1}{c|}{\multirow{4}{*}{$0$}}         & \multicolumn{1}{c|}{Model 0}                   & \multicolumn{1}{c|}{$-0.11 \pm 0.31$}  & \multicolumn{1}{c|}{$-0.03 \pm 0.29$}  & $-0.01 \pm 0.29$  \\
\multicolumn{1}{c|}{}                           & \multicolumn{1}{c|}{Model 1}                    & \multicolumn{1}{c|}{$-0.13 \pm 0.34$}  & \multicolumn{1}{c|}{$-0.00 \pm 0.32$}  & $0.02 \pm 0.33$   \\
\multicolumn{1}{c|}{}                           & \multicolumn{1}{c|}{\multirow{2}{*}{Model 2}} & \multicolumn{1}{c|}{\textcolor{BrickRed}{$0.56\pm 0.35$}}    & \multicolumn{1}{c|}{\textcolor{BrickRed}{$0.88\pm 0.55$}}    & \textcolor{BrickRed}{$0.71 \pm 0.37$}   \\
\multicolumn{1}{c|}{}                           & \multicolumn{1}{c|}{}                              & \multicolumn{1}{c|}{($0.05 \pm 0.54$)} & \multicolumn{1}{c|}{($0.07 \pm 0.71$)} & $(0.06 \pm 1.19)$   \\ \hline\hline
\multicolumn{1}{c|}{\multirow{4}{*}{$3$}}         & \multicolumn{1}{c|}{Model 0}                   & \multicolumn{1}{c|}{$3.08 \pm 0.57$}   & \multicolumn{1}{c|}{$2.99 \pm 0.52$}   & $3.01 \pm 0.53$   \\
\multicolumn{1}{c|}{}                           & \multicolumn{1}{c|}{Model 1}                    & \multicolumn{1}{c|}{$3.08 \pm 0.58$}   & \multicolumn{1}{c|}{$3.05 \pm 0.55$}   & $3.04 \pm 0.55$   \\
\multicolumn{1}{c|}{}                           & \multicolumn{1}{c|}{\multirow{2}{*}{Model 2}} & \multicolumn{1}{c|}{\textcolor{BrickRed}{$3.71 \pm 0.67$}}   & \multicolumn{1}{c|}{\textcolor{BrickRed}{$3.92 \pm 0.73$}}   & \textcolor{BrickRed}{$3.68 \pm 0.66$}   \\
\multicolumn{1}{c|}{}                           & \multicolumn{1}{c|}{}                              & \multicolumn{1}{c|}{($3.23 \pm 0.77$)} & \multicolumn{1}{c|}{($3.09 \pm 0.90$)} & ($3.09 \pm 1.75$) \\
\bottomrule
\end{tabular}
\caption{Estimated mean $\bar{r}$ and standard deviation $\sigma(r)$ of the inferred tensor-to-scalar ratio $r$, in units of $10^{-3}$. The results are obtained from the three foreground simulation suites (Models 0, 1, and 2), using three component-separation methods: Pipeline A, B, and C. Results are provided for simulations with input $r = 0$ and $r = 0.003$ (for a total of 500 simulations). Numbers in red highlight configurations for which the resulting bias on $r$ is greater than $1\sigma$. Results obtained from the extended pipelines are shown only for Model 2 in parentheses. With the model extension, all three pipelines are able to recover $r$ without bias.}
\label{tab:r_bias_unc}
\end{table}

To counteract the biases observed in foreground Model~2, we modify the nominal pipelines to include additional levels of foreground marginalization, as outlined in Sec.~\ref{sec:likelihood}.
Specifically, for Pipeline A we allow the dust and synchrotron decorrelation parameters $\Delta'_{\rm d}$ and $\Delta'_{\rm s}$ to vary.\footnote{We assume here that the (frequency) decorrelation parameters $\Delta_{\rm d}$ and $\Delta_{\rm s}$ do not depend on the angular scale, i.e., we assume it is flat in $\ell$ space.}
For the map-level cleaning methods, we add additional parameters that capture the effect of the residual contamination in the cleaned CMB maps. 
In Pipeline B we propagate a power-law dust spectrum through the component-separation weights and marginalize over its amplitude $A_{\rm d}$ and spectral slope $\alpha_{\rm d}$ (see Eq.~\ref{eq:fg_marg_ilc}), while for pipeline C we marginalize over the amplitude $A'_{\rm d}/A'_{\rm s}$ of the $BB$ spectra of the component-separated dust and synchrotron maps (see Eq.~\ref{eq:fg_marg_ml}).
These modifications reduce the foreground-induced biases to $\lesssim 0.1\sigma$ for all pipelines (see Fig.~\ref{fig:r_bias_unc} and Tab.~\ref{tab:r_bias_unc}). 
However, these changes also lead to increased statistical uncertainties due to the enlarged parameter space and potential parameter degeneracy. 
The $\sigma(r)$ degradation varies by method, increasing by 29\% for Pipeline B, 54\% for Pipeline A, and up to 100\% for Pipeline C.
We also assess the relative impact of dust and synchrotron contributions to the bias on $r$ in Model 3 by rerunning the analysis for pipeline A for the $r=0$ case, allowing decorrelation only for dust. This reduces the bias from $r [\times 10^3] = 0.56 \pm 0.35$ to $r [\times 10^3] = 0.03 \pm 0.48$, suggesting that dust is the dominant contaminant.\\
\begin{figure*}
    \centering
    \includegraphics[width=0.9\textwidth]{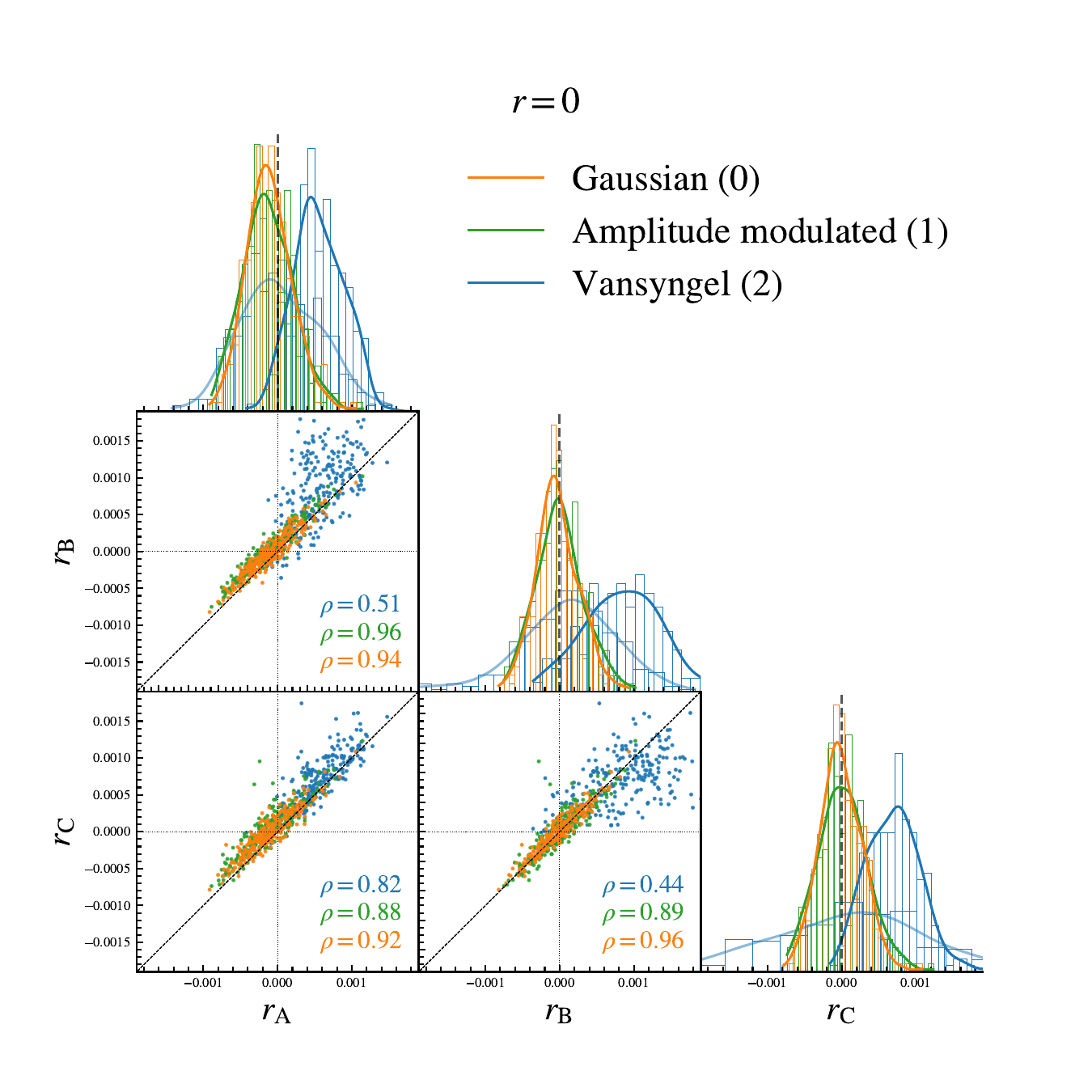}
    \caption{Scatter plots and histograms displaying best-fit tensor-to-scalar ratio values from different component-separation pipelines for three foreground models in the $r=0$ scenario. Each panel provides pairwise comparisons of the pipelines. The scatter plots in the off-diagonal panels show correlations of the $r$-values between the pipelines, with Pearson correlation coefficients indicated by $\rho$. The histograms on the diagonal display the distribution of maximum-likelihood estimates of $r$ values for each pipeline, with smooth lines representing kernel density estimates (KDE) of the histograms. Colors orange, green, and blue represent foreground models 1, 2, and 3, respectively. The faint blue line in the histograms illustrates the case of extended foreground marginalization applied when considering foreground model 2.}
    \label{fig:scatter_plot_r}
\end{figure*}
%

For simulations that include a PGW component of $r=0.003$ we measure $\sigma(r)$ to fall between $5.2  \times 10^{-4}$ and $5.8 \times 10^{-4}$ in both the Gaussian and amplitude-modulated foreground models, corresponding to a detection of primordial tensor power at approximately $5\sigma$ for $r=0.003$. 
In scenarios involving foreground model 2, the nominal pipelines exhibit a bias close to $1\sigma$. 
However, with the implementation of additional foreground marginalization, the pipelines yield mean recovered $r$ values that align with the input $r=0.003$, within statistical uncertainties, ranging from 0.05$\sigma$ (Pipeline C) to 0.3$\sigma$ Pipeline A). 
Nevertheless, this extension exacerbates the statistical uncertainty $\sigma(r)$ across all pipelines, increasing by 15--20\% for Pipelines A and B, and by as much as 160\% for Pipeline C. 
This translates to a 3.4$\sigma$ and $4.2\sigma$ detection of PGWs with $r=0.003$ using Pipeline A and B, but for Pipeline C the significance of the measurement decreases to $1.8\sigma$.\\

All the constraints on the tensor-to-scalar ratio quoted so far include the impact of delensing. 
To get a sense of the lensing contribution to the statistical uncertainty on $r$, we repeat the analysis on simulations without the delensing step.
We find that delensing typically reduces $\sigma(r)$ by a factor $\approx 10$ and $\approx 6$ for $r=0$ and $r=0.003$ simulations, respectively.\\

Assessing the relative performance of different foreground-cleaning methods is not a straightforward task given the different assumptions on the data model and map processing that each pipeline makes.
To provide deeper insights into this comparison, Fig.~\ref{fig:scatter_plot_r} presents a scatter plot of the tensor-to-scalar ratios estimated from 250 realizations across different foreground models and component-separation algorithms. 
Pipelines A, B, and C show high correlation in inferring $r$ values when dealing with simpler foreground models such as Models 0 and 1. 
Correlation coefficients are notably high (above 0.90 in most cases), indicating that all methods are consistent with each other in these scenarios.
For foreground model 2, which presents more complex and potentially closer to realistic foreground conditions, the correlations are notably lower between parametric (A and C) and non-parametric (B) methods. 
For instance, the correlation between Pipeline B and C drops to as low as 0.44, and with Pipeline A to 0.51.
In App.~\ref{sec:corr_estimator} we provide a summary of the Pearson correlation coefficients between different pipelines for various foreground suites, for the $r=0.003$ scenario, and a discussion of their expectation values.
When primordial tensor power is present in the simulations, i.e., $r=0.003$, we find the correlation coefficients to be comparatively larger due to the presence of the CMB signal.
We note that the correlation of Pipeline C with the other methods drops significantly after marginalizing over residual foregrounds. \\

Although the primary focus of this paper is the comparison of different foreground cleaning methods to recover the tensor-to-scalar ratio, we point the interested reader to App.~\ref{sec:foreground_results} for additional insights into what parametric methods can reveal about foreground emission.

\subsection{Application to \texttt{PySM} models and MCMC results}  
\label{sec:results_pysm}  

CMB-S4 will provide us with a single multi-frequency measurement of the microwave sky. 
Therefore we will have to determine consistency between our different pipelines given a single realization. 
In this subsection we will present and analyze results based on a single realization of the sky. 
For this task, we use the \texttt{PySM3} foreground suites introduced in Sec.~\ref{sec:pysm_models}. 
The three \texttt{PySM3} models chosen for this analysis (Models 3, 4, and 5) each have unique assumptions about the physical properties and spatial distributions of foregrounds such as Galactic dust and synchrotron emission. 
Most relevantly for our three cleaning methods, the low complexity suite (Model 3) is free of frequency decorrelation, while the medium complexity suite (Model 4) incorporates frequency decorrelation via spatially varying spectral parameters of the dust and synchrotron SED. 
The high complexity suite (Model 5) incorporates a 3D model of the polarized dust emission, causing more complicated decorrelation effects.
As with the Vansyngel model (Model 2), we have only a single realization of the foreground emission for each of Models 3, 4, and 5. 
However, unlike the previous section, where we examined the maximum-likelihood $r$ values across 500 simulations, here we use MCMC to analyze just one mock sky. 
Importantly, the simulations for Models 3, 4, and 5 all share the same noise realization, enabling a direct comparison of posterior results on equal footing.\\

To sample the full posterior distribution, we use a Markov chain Monte Carlo (MCMC) method, specifically the Metropolis-Hastings algorithm with adaptive covariance learning, as implemented in the publicly-available \texttt{Cobaya} package \citep{cobaya}. 
\\

\begin{figure}
    \centering
    \includegraphics[width=\columnwidth]{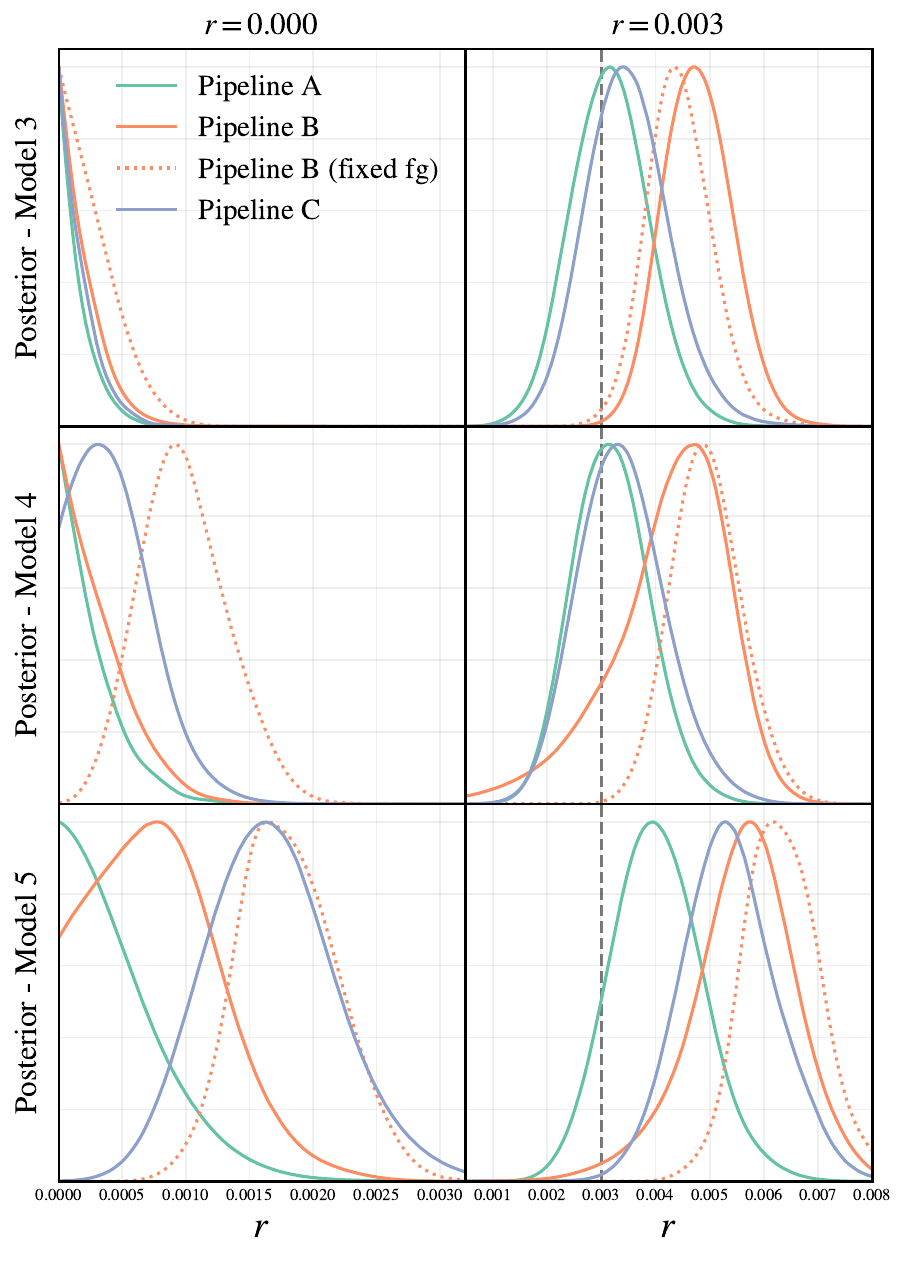}
    \caption{Posterior distributions of the tensor-to-scalar ratio ($r$) derived from MCMC chains run on \texttt{PySM3} foreground models. The rows of panels, from top to bottom, display results for Models 3, 4, and 5. The left column corresponds to the $r=0$ scenario, while the right column shows results for $r=0.003$. In each panel, the colored lines represent the posterior distributions obtained from the three analysis pipelines (A, B, and C). The dotted line represents Pipeline B results without marginalizing over residual foregrounds, typically leading to a larger bias in $r$. The vertical black dashed line marks the input value of $r=0.003$ for reference.}
    \label{fig:posterior_r_pysm}
\end{figure}

In Fig.~\ref{fig:posterior_r_pysm}, we present the posterior distributions of $r$ obtained from the MCMC chains. 
Rows correspond to one of the three \texttt{PySM3} foreground models (Models 3, 4, and 5).
The left column displays the posterior distributions for the $r=0$ case, while the right column shows the results for $r=0.003$. 
Within each panel, we compare the performance of pipelines A, B, and C, with the different colors representing the posterior distributions for each pipeline.
To remain agnostic about the complexity of different foreground realizations, all pipeline implementations incorporate additional marginalization over residual foregrounds, as described in Sec.~\ref{sec:likelihood}.
The inferred 2$\sigma$ credibility intervals on the tensor-to-scalar ratio are also summarized in Tab.~\ref{tab:r_pysm}.\footnote{The LT pipeline, developed in \citet{belkner2023cmbs4}, was undergoing refinement when we began the analysis on Models 0, 1, and 2. By the time we analyzed these \texttt{PySM} models, the pipeline had been significantly improved, with a ~5\% increase in delensing efficiency. Re-analysis of Model 0 using the updated pipeline yielded slightly tighter constraints on $r$, emphasizing the critical role of delensing for accurate primordial $B$-mode estimation.}
\\

Pipeline A successfully recovers the marginal posterior distribution of $r$, remaining consistent with the input values to within $1\sigma$ across all scenarios.
In contrast, Pipeline B shows a slight positive bias for non-zero input values of $r$, although it remains consistent with $r=0$ in cases of low and medium foreground complexity (Models 3 and 4). 
As illustrated by the dotted red lines in Fig.~\ref{fig:posterior_r_pysm}, the posteriors on $r$ obtained without marginalizing over residual foregrounds in the ILC spectra exhibit a more pronounced bias, underscoring the importance of accounting for foreground residuals. 
While Pipeline C produces $r$ estimates that are consistent with the input for low and medium complexity models, Pipeline C does not recover an unbiased posterior for the high complexity models.
These results suggest an insufficient modeling of the foreground residual caused by foreground emission anisotropy and spatial variation of the foreground SED for Pipelines B and C.
Potential improvements could include using a spatially varying mixing matrix, $\beta_{\rm d/s} = \beta_{\rm d/s}(\nver)$, in Pipeline C, either through a multi-patch approach \citep{Errard:2018ctl} or clustering methods \citep{Puglisi:2021hqe}, or operating in the needlet domain to adapt weights for local variations of the foreground emission in Pipeline B \citep{Delabrouille_2008}.\\

These results highlight the importance of employing a multifaceted approach to $r$ inference, wherein multiple pipelines are utilized and consistency across methods is evaluated. 

\begin{table}[t]
\renewcommand{\arraystretch}{1.5} 
\begin{tabular}{ccccc}
\toprule
\rowcolor{gray!30} \multicolumn{5}{c}{95\% credible intervals on $r$ $[\times 10^3]$}                                                                                                                                               \\ \hline
\hline
\multicolumn{1}{c|}{\multirow{1}{*}{Input $r$}} & \multicolumn{1}{c|}{\multirow{1}{*}{FG model}}     & \multicolumn{1}{c|}{Pipeline A}   & \multicolumn{1}{c|}{Pipeline B}               & Pipeline C                \\ \hline
\multicolumn{1}{c|}{\multirow{3}{*}{$0$}}         & \multicolumn{1}{c|}{Model 3}                   & \multicolumn{1}{c|}{$< 0.44$}  & \multicolumn{1}{c|}{$< 0.54$}  & $<0.47$  \\
\multicolumn{1}{c|}{}                           & \multicolumn{1}{c|}{Model 4}                    & \multicolumn{1}{c|}{$< 0.75$}  & \multicolumn{1}{c|}{$< 0.86$}  & $< 1.0$   \\
\multicolumn{1}{c|}{}                           & \multicolumn{1}{c|}{Model 5}                              & \multicolumn{1}{c|}{$<1.3$} & \multicolumn{1}{c|}{$<1.7$} & $1.7 \pm 1.0 $   \\ \hline\hline
\multicolumn{1}{c|}{\multirow{3}{*}{$3$}}         & \multicolumn{1}{c|}{Model 3}                   & \multicolumn{1}{c|}{$3.2\pm 1.4$}   & \multicolumn{1}{c|}{$4.8^{+1.3}_{-1.2}$}   & $3.5^{+1.6}_{-1.4}$   \\
\multicolumn{1}{c|}{}                           & \multicolumn{1}{c|}{Model 4}                    & \multicolumn{1}{c|}{$3.2\pm 1.4$}   & \multicolumn{1}{c|}{$4.2^{+1.9}_{-2.4}$}   & $3.4^{+1.6}_{-1.5}$   \\
\multicolumn{1}{c|}{}                           & \multicolumn{1}{c|}{Model 5}                              & \multicolumn{1}{c|}{$4.0 \pm 1.7$} & \multicolumn{1}{c|}{$5.6^{+1.9}_{-2.0}$} & $5.4\pm 1.7$ \\
\bottomrule
\end{tabular}
\caption{$2\sigma$ credibility intervals inferred from the analysis of the \texttt{PySM}-based foreground suites.}
\label{tab:r_pysm}
\end{table}


\section{Conclusions} 
\label{sec:conclusions}
This paper presents a thorough comparative analysis of a few proposed component-separation methods for inflationary gravitational wave searches using large-scale $B$-mode polarization measurements from the next-generation CMB-S4 experiment’s Small Aperture Telescopes (SATs). 
We evaluated three distinct foreground cleaning strategies: a parametric $C_\ell$-based method (Pipeline A); a non-parametric map-based method (Pipeline B; ILC); and a parametric map-based method (Pipeline C). Additionally, we explored extensions of these nominal cleaning methods to marginalize residual foreground contamination. 
For the $C_\ell$-based pipeline, we allowed dust and synchrotron decorrelation parameters to vary freely, while map-based pipelines included spectral templates for dust and synchrotron power. 
Although not the primary focus, our pipelines also incorporated realistic SAT delensing through a $B$-mode lensing template, constructed using an optimal lensing reconstruction scheme applied to mock maps from a fiducial CMB-S4 Large Aperture Telescope (see Ref.~\citep{belkner2023cmbs4} for further details).\\

We validated the performance of these methods on realistic simulations of the microwave sky, including lensed CMB polarization anisotropies with and without primordial tensor modes, instrumental noise, and three foreground models with varying degrees of complexity.
When considering nominal pipelines, we found that the inferred sensitivity to the tensor-to-scalar ratio is largely comparable across the three cleaning methods, and they range from $\sigma(r) = 3 \times 10^{-4}$ to $\sigma(r) = 5 \times 10^{-4}$ (for $r=0$).
While all pipelines recovered unbiased values for $r$ in the Gaussian and amplitude-modulated simulations, biases at the $\gtrsim 1\sigma$ level appeared when analyzing the more complex Vansyngel foreground suite, which features spatial variations in the dust spectral index and frequency decorrelation. 
We were able to recover unbiased $r$ when using extended pipelines that marginalized over frequency decorrelation and foreground residual power, though with increased $\sigma(r)$—15–20\% for Pipelines A and B, and up to 160\% for Pipeline C.
This translates to a detection significance between $2\sigma$ and $4\sigma$ for an expected input value of $r=0.003$. We applied the same methods to more recent \texttt{PySM} foreground models, showing that while the frequency decorrelation model in Pipeline A remains robust, further work is required to model foreground residual power in Pipelines B and C.\\

This work is one of the first to address foreground biases across these methods using the same set of simulations at these low-noise levels with realistic delensing included. 
Future work will address additional data complexities, including anisotropic noise and time-ordered data filtering, as well as the impact of systematic effects such as calibration errors and bandpass uncertainties, e.g., Ref.~\citep{dick10}. 
As of early 2025, CMB-S4 is assessing the feasibility of $r$ measurements from a Chilean site, where foreground challenges are expected to be more significant. 
Although this study focused on the South Pole Deep Patch, the lessons gleaned for component separation
at these ultra-low noise levels will provide an instructive foundation for 
the more demanding foreground conditions in the new observational context.

\begin{acknowledgments}
We would like to thank Carlo Baccigalupi, Aurelien Fraisse, Carlos Hervías-Caimapo, Sunguen Kim, Bruce Partridge, Giuseppe Puglisi, Douglas Scott, Ahmed Soliman, Raked Stompor, Matthieu Tristram, Sarvesh Kumar Yadav, and members of the CMB-S4 low-ell BB analysis working group for providing comments and feedback to this work. 
CMB-S4 is supported by the Director, Office of Science, Office of High Energy Physics of the U.S. Department of Energy under contract No. DE- AC02-05CH11231; by the National Energy Research Scientific Computing Center, a DOE Office of Science User Facility under the same contract; and by the Divisions of Physics and Astronomical Sciences and the Office of Polar Programs of the U.S. National Science Foundation under Mid-Scale Research Infrastructure award OPP-1935892.
This work used the resources of the SLAC Shared Science Data Facility (S3DF) at SLAC National Accelerator Laboratory. S3DF is a shared High-Performance Computing facility, operated by SLAC, that supports the scientific and data-intensive computing needs of all experimental facilities and programs of the SLAC National Accelerator Laboratory. 
The SLAC authors acknowledge support by the Department of Energy, Contract DE-AC02-76SF00515.
W.L.K.W is supported in part by the Department of Energy, Laboratory Directed Research and Development program and as part of the Panofsky Fellowship program at SLAC National Accelerator Laboratory.
The research was carried out in part at the Jet Propulsion Laboratory, California Institute of Technology, under a contract with the National Aeronautics and Space Administration (80NM0018D0004).
\end{acknowledgments}
 
\bibliographystyle{apsrev4-2} 
\bibliography{apssamp}

\appendix
\section{Auto-and Cross-$BB$ Spectra from the CMB-S4 South Pole Deep Patch}
\label{app:s4_bb_crossfreq}
\begin{figure*}
   \centering
   \includegraphics[width=\textwidth]{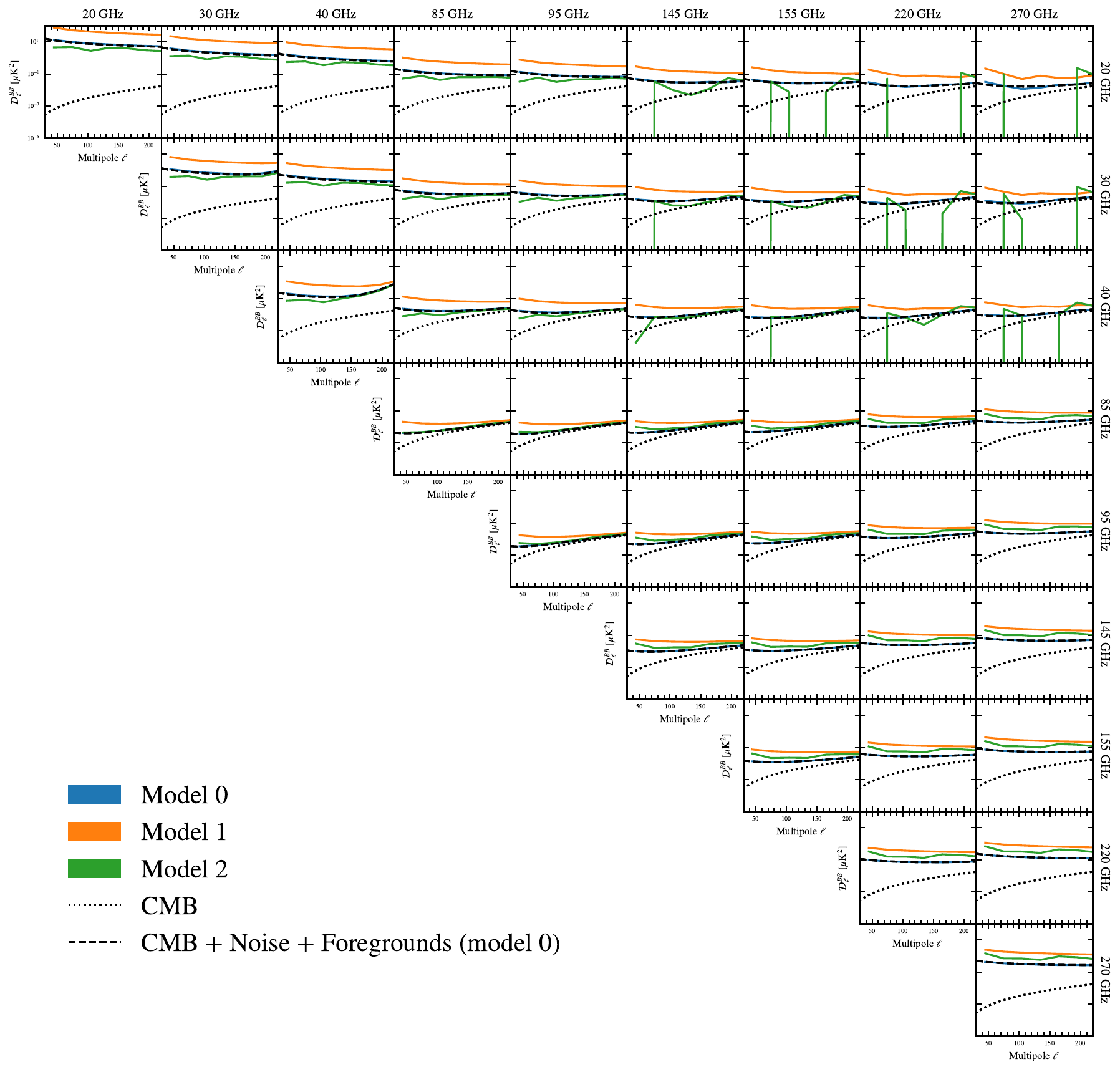}
   \caption{Mean set of auto-and cross-$BB$ spectra from the CMB-S4 South Pole deep patch across different frequencies and foreground models. The three different models are Model 0 (Gaussian simulations, shown in blue), Model 1 (amplitude-modulated simulations, shown in orange), and Model 2 (Vansyngel simulations, shown in green). The auto-spectra are not noise debiased. The dotted black line represents the lensed CMB spectrum, while the dashed black line represents the combination of the lensed CMB, foreground components (from Model 0), and instrumental noise (for auto-spectra). The blue lines completely overlap with the black dashed lines in all cases.}
   \label{fig:s4_bb_crossfreq}
\end{figure*}

In this section, we present the mean set of auto-and cross-$BB$ spectra obtained from the CMB-S4 South Pole deep patch. The spectra are shown across different frequencies and foreground models, providing a breakdown of the polarization data. Fig.~\ref{fig:s4_bb_crossfreq} illustrates the mean spectra from three different simulation models: Gaussian simulations; amplitude-modulated simulations; and Vansyngel simulations.\\

Note that the mean auto-spectra shown in Fig.~\ref{fig:s4_bb_crossfreq} are not noise-debiased. 
The dotted black line represents the lensed CMB spectrum, indicating the expected signal from the gravitational lensing of the CMB. 
The dashed black line shows the combined effect of the lensed CMB, foreground components (from Model 0), and instrumental noise (for auto-spectra). 
Interestingly, the spatial-dependent modulation of the foreground brightness in Model 1 results in an overall higher amplitude of the $BB$ power spectra computed between different frequencies.
This breakdown showcases the contributions  of different components in the observed cross-frequency $BB$ spectra and aids in interpreting the foreground parameter fits in Pipeline A.\\

\section{Lensing template validation}
\label{app:lt_validation}
\begin{figure}
    \includegraphics[width=\columnwidth]{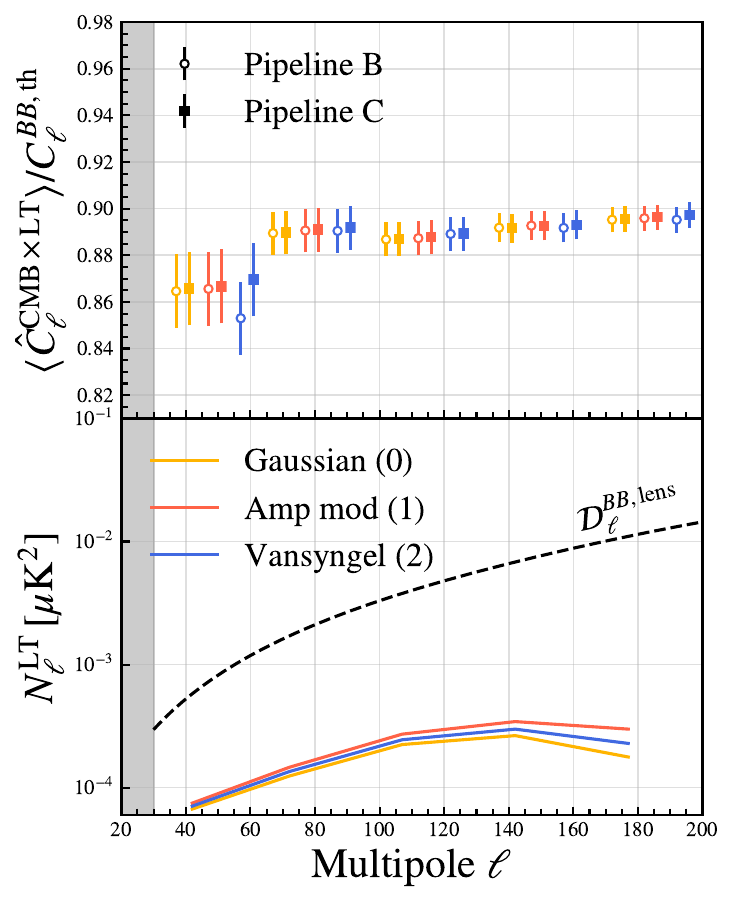}
    \caption{\textit{Top panel}: Average cross-power spectrum between the cleaned CMB maps from Pipelines B (empty circles) and C (filled squares) and the lensing template. Different colors represent the results for the three foreground suites. The spectra are normalized to the input lensed $BB$ power spectrum, and the error bars are divided by $\sqrt{N_{\rm sims}}$. \textit{Bottom panel}: Lensing template noise power spectrum, calculated using Eq.~\ref{eqn:ltnoise}, with the same color coding as in the top panel. The black dashed line represents the fiducial lensing $B$-mode power.
The auto-spectrum of the lensing template and the cross-spectrum between the lensing template and the cleaned CMB maps form part of the input data vector for the likelihood. This illustrates the high degree of recovery in the lensing $B$ modes, consistent with Ref.~\cite{belkner2023cmbs4}, and highlights the agreement in cross-power and uncertainties between Pipelines B and C across different foreground models.}
    \label{fig:lt_plots}
\end{figure}

As an additional validation of the lensing template construction, in the top panel of Fig.~\ref{fig:lt_plots}, we show the average cross-correlation between the cleaned CMB maps obtained from both Pipelines B and C (empty circles and filled squares respectively) and the LT across our three foreground models (different colors).
Note that the spectra are normalized to the input lensed $BB$ power spectrum.
The bottom panel shows the estimated LT noise power spectrum for the three foreground suites using Eq.~\ref{eqn:ltnoise}, which are used to noise-debias the lensing template auto-spectrum.\\

\begin{figure}
    \includegraphics[width=\columnwidth]{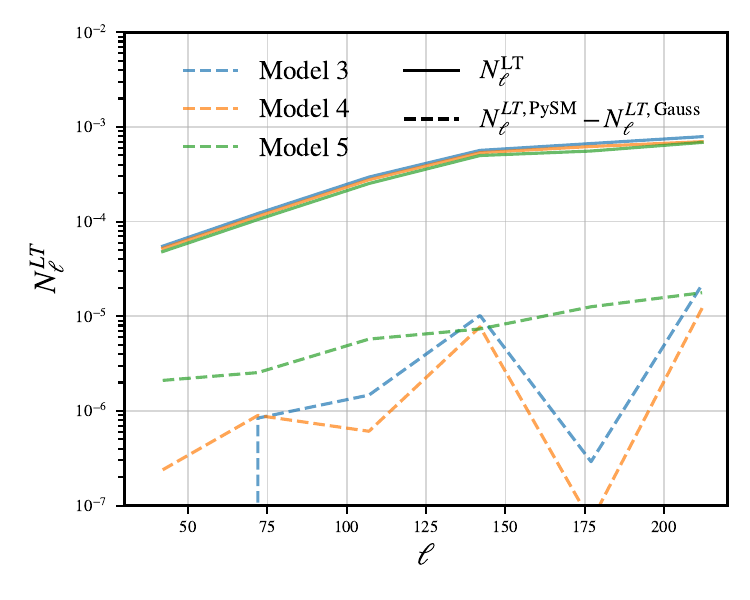}
    \caption{Lensing template noise bias for foreground models 3--5 (see Sec.~\ref{sec:pysm_models}). We show the difference between the idealistic estimate of the noise bias including full knowledge of the respective foreground template ($N^\mathrm{LT,PySM}$) and the more realistic estimate of a fully Gaussian foreground simulation with matching two-point power ($N^\mathrm{LT,Gauss}$). In a realistic analysis only the latter could be used and the potential bias due to the non-Gaussian/higher-order correlations are not captured in these simulations as shown in the dashed lines.}
    \label{fig:lt_ng_plots}
\end{figure}

In our analysis, we estimate the LT noise
from CMB-S4 LAT simulations, which include the respective foreground suite's sky templates.
In debiasing the LT auto-spectrum with this $N_\ell^{\rm LT}$, we rely on perfect knowledge of the foregrounds for this producure to be unbiased.
However, this cannot be achieved in real data at the same accuracy.
While the residual foreground level in the ILC-cleaned LAT maps is relatively low and the foreground bias in the reconstructed lensing template is subdominant, higher-order correlations of the foregrounds could still cause biases in the lensing template auto-power spectrum \cite{Beck2020}. 
To assess the amplitude of the higher-order biases in the LT auto-spectrum from the beyond-Gaussian power of the foregrounds, we construct a parallel pipeline to estimate $N_\ell^{\rm LT}$. 
In this pipeline, we replace the LAT foregrounds with Gaussian realizations:
we fit a simple foreground model to the LAT multi-frequency maps using pipeline A, generate a set of 100 Gaussian simulations from the resulting best-fit foreground power spectra for each of the three \texttt{PySM} models, and run the LT-generation pipeline on this new set of foreground simulations. 
The mean auto-power spectrum of these simulations can be used as an estimate for $N^\mathrm{LT}$. We show the difference between this estimate ($N^\mathrm{LT,Gauss}$) and the idealistic estimate of using the true foreground template in the simulations ($N^\mathrm{LT,PySM}$) in Fig.~\ref{fig:lt_ng_plots}. 
The incurred bias due to not including higher-order foreground correlations is below 2\%.

We conclude that the difference between $N^\mathrm{LT,PySM}$ and $N^\mathrm{LT,Gauss}$ is negligible for the foreground simulations in our patch, and thus proceed with using $N^\mathrm{LT,PySM}$ templates to debias the LT auto-spectrum in this work. 
However, we note that for real data analyses, we do not have a perfect map of foregrounds to construct the analogous term to $N^\mathrm{LT,PySM}$ and the approach of using $N^\mathrm{LT, Gauss}$ can be adopted.

\section{Correlation between $r$ estimators}
\label{sec:corr_estimator}
\begin{figure*}
    \centering
    \includegraphics[width=\textwidth]{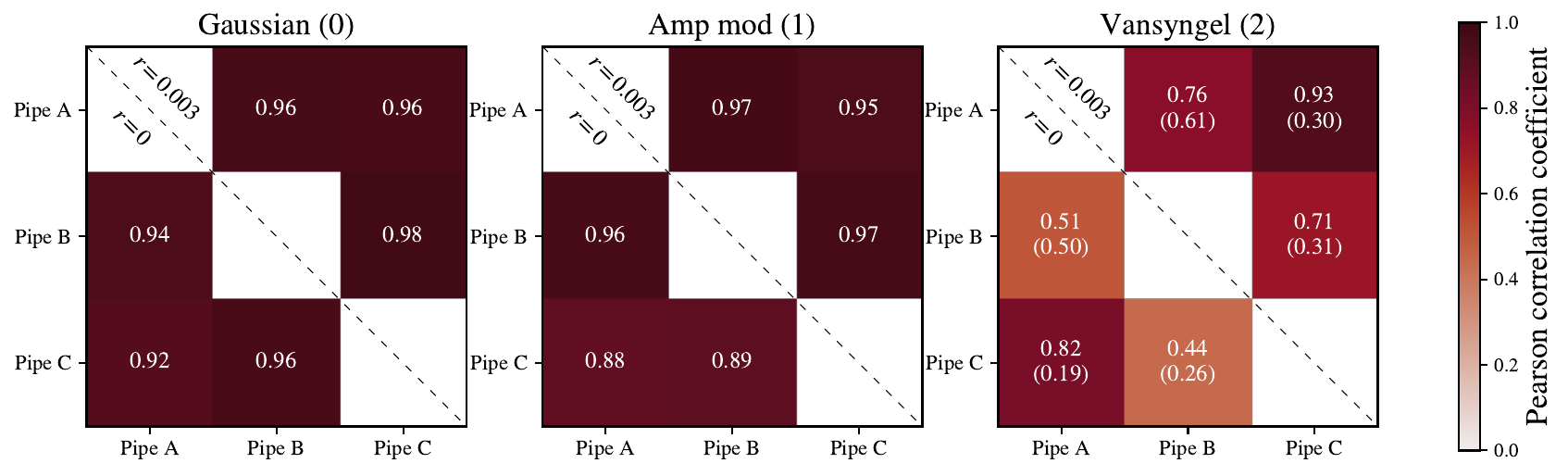}
    \caption{Correlation matrices for the three foreground simulation models. Each matrix evaluates the Pearson correlation coefficients calculated from the inferred tensor-to-scalar ratios $r$ using the three distinct foreground cleaning methods on 250 simulations. In each panel, the upper triangle matrix shows results for $r=0.003$ while the elements in the lower triangle matrix correspond to $r=0$. The matrices are color-coded from 0 to 1. For foreground model 2, effects of extending the nominal pipelines to marginalize over residual foreground contamination are noted in parentheses.}
    \label{fig:corr_mats_r}
\end{figure*}
\begin{figure*}
    \centering
    \includegraphics[width=\textwidth]{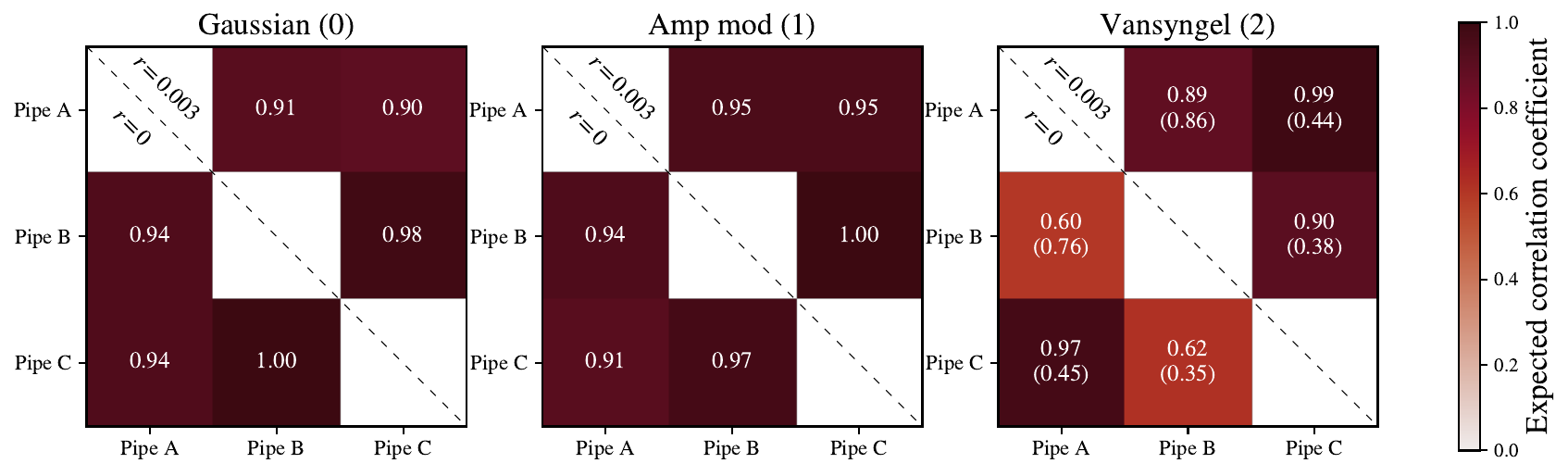}
    \caption{Same figure as Fig.~\ref{fig:corr_mats_r}, but predicting the correlation coefficients from the spread of the distributions of $r$ summarized in Tab.~\ref{tab:r_bias_unc} for each foreground model.}
    \label{fig:corr_mats_r_expected}
\end{figure*}
Determining the optimal $r$ estimator for CMB-S4 remains an open question. 
As shown in Sec.~\ref{sec:results}, each pipeline exhibits sensitivity to different signal and noise modes, leading to imperfect correlations between the estimators. 
This appendix quantifies and models these empirical correlations.

Fig.~\ref{fig:corr_mats_r} shows the Pearson correlation coefficients between the pipelines for various foreground suites, including the $r = 0.003$ case. 
As expected, correlations are generally higher when primordial tensor power is present ($r = 0.003$) due to the shared CMB signal. 
Notably, the correlation of Pipeline C with the other methods decreases after marginalizing over residual foregrounds.\\

To understand these correlations, we consider the result of a maximum likelihood estimator having a limiting normal distribution \cite{esl},
$$
\hat{r}^i \rightarrow \mathcal{N}(\mathbf{r}_0,\mathbf{\Sigma}),
$$
where $\mathcal{N}$ is a multivariate Gaussian distribution function and $\mathbf{\Sigma}$ is the Fisher information matrix. In our case we consider three statistical variables $i=A,B,C$ corresponding to the three estimates of $r$ from three different pipelines run on the same data. We model the Fisher information matrix as
$$
\mathbf{\Sigma}=
\begin{pmatrix}
\sigma^2+\Delta\sigma_A^2 & \sigma^2 & \sigma^2\\
\sigma^2 & \sigma^2+\Delta\sigma_B^2 & \sigma^2\\
\sigma^2 & \sigma^2 & \sigma^2+\Delta\sigma_C^2
\end{pmatrix},
$$
where $\sigma^2$ represents the variance of $r$ due to CMB and instrumental noise, and $\Delta\sigma^2_i$ is the estimator-specific variance arising from factors like foreground residuals, noise-weighting differences, or numerical noise. 
This model yields cross-correlation coefficient:
$$
\rho_{ij}=\frac{1}{\sqrt{\left(1+\frac{\Delta\sigma_i^2}{\sigma^2}\right)\left(1+\frac{\Delta\sigma_j^2}{\sigma^2}\right)}} \ \textrm{for }i\neq j.
$$
This equation connects the measured $\sigma(r)$ values reported in the previous Sec.~\ref{sec:r_constraints} to the inter-pipeline correlation coefficients. \\

We calculate $\sigma$ using a minimum-variance combination of the three pipelines based on the simulation-inferred covariance matrix. 
The corresponding $\Delta\sigma$ values are then derived from each pipeline's distribution of best-fit $r$ values.
Fig.~\ref{fig:corr_mats_r_expected} shows the resulting predicted correlation coefficients, which agree reasonably with the empirical correlations in Fig.~\ref{fig:corr_mats_r}. 
This agreement suggests that the decorrelation between $r$ estimates from different pipelines is consistent with the differences in their respective $\sigma(r)$ values.

\section{Recovered foreground analysis \label{sec:foreground_results}}
\begin{figure*}
    \centering
    \includegraphics[width=0.96\textwidth]{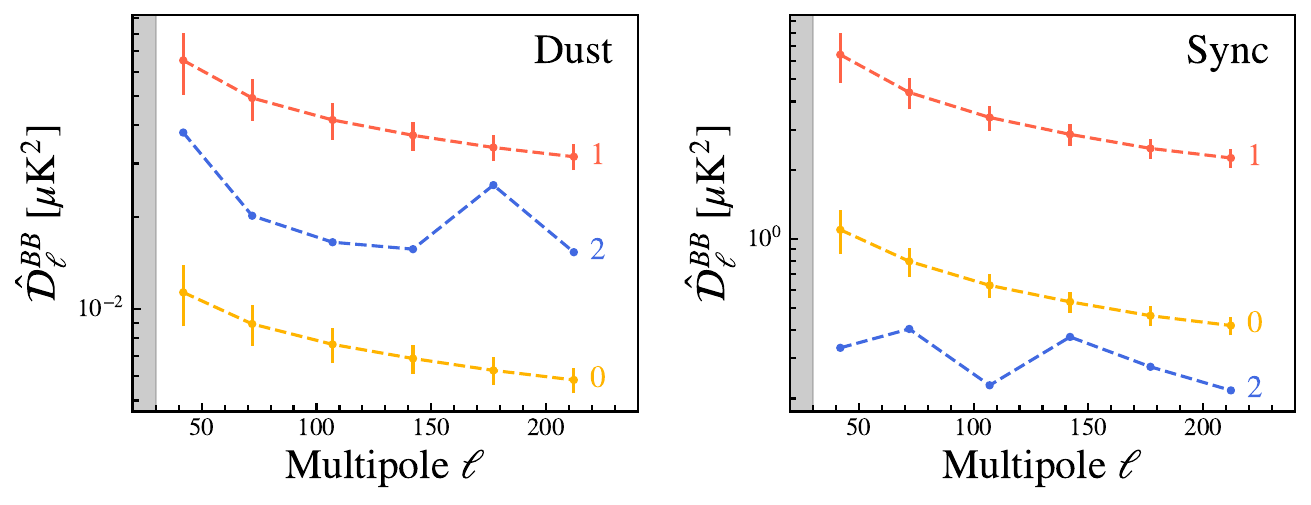}
    \caption{Mean foreground spectra recovered by the parametric map-based maximum likelihood Pipeline C. The left panel shows the mean Galactic dust spectra, while the right panel displays the mean synchrotron spectra. Different colors (orange, red, and blue) in each panel represent results from the various foreground suites (0, 1, and 2). The error bars correspond to a single realization. The uncertainties for Model 2 appear smaller compared to other models, as only one full-sky realization of the foreground power is available. The spectra have been debiased for instrumental noise.}
    \label{fig:dc0800vs07vs09_ML_spectra_dust_sync}
\end{figure*}

In this appendix, we explore the foreground information extracted by parametric component separation methods. As opposed to Pipelines A and B, we obtain separate CMB, dust and synchrotron component maps in Pipeline C. 
These maps can be used to further study foreground components that could feed back into improving our component-separation methods.
In Fig.~\ref{fig:dc0800vs07vs09_ML_spectra_dust_sync}, we present the average $BB$ spectra of Galactic dust and synchrotron as recovered by Pipeline C across the different foreground models, depicted with orange, red, and blue curves, respectively. 
Dust maps are output at a reference frequency of 353\,GHz, while synchrotron maps are at 23\,GHz. 
The Gaussian foreground model confirms the efficacy of Pipeline C, as the recovered spectra align with the power-law power spectrum used as input in the simulation generation. 
The amplitude-modulated Model 1 shows a power-law spectrum with increased amplitude compared to the fully Gaussian Model 0. 
In contrast, the dust and synchrotron templates derived from the more complex and data-driven Vansyngel Model 2 exhibits clear deviations from the power-law spectrum shape. 
We observe a large dust amplitude and a relatively low amplitude for synchrotron, consistent with the parametric estimates of $A_{\rm d}$ and $A_{\rm s}$ in Pipeline A. 
These recovered $BB$ power spectra serve as foreground templates in estimating $r$ with Pipeline C maps for extended foreground marginalization, as detailed in Sec.~\ref{sec:likelihood}.\\

\begin{figure*}
    \centering
    \includegraphics[width=\textwidth]{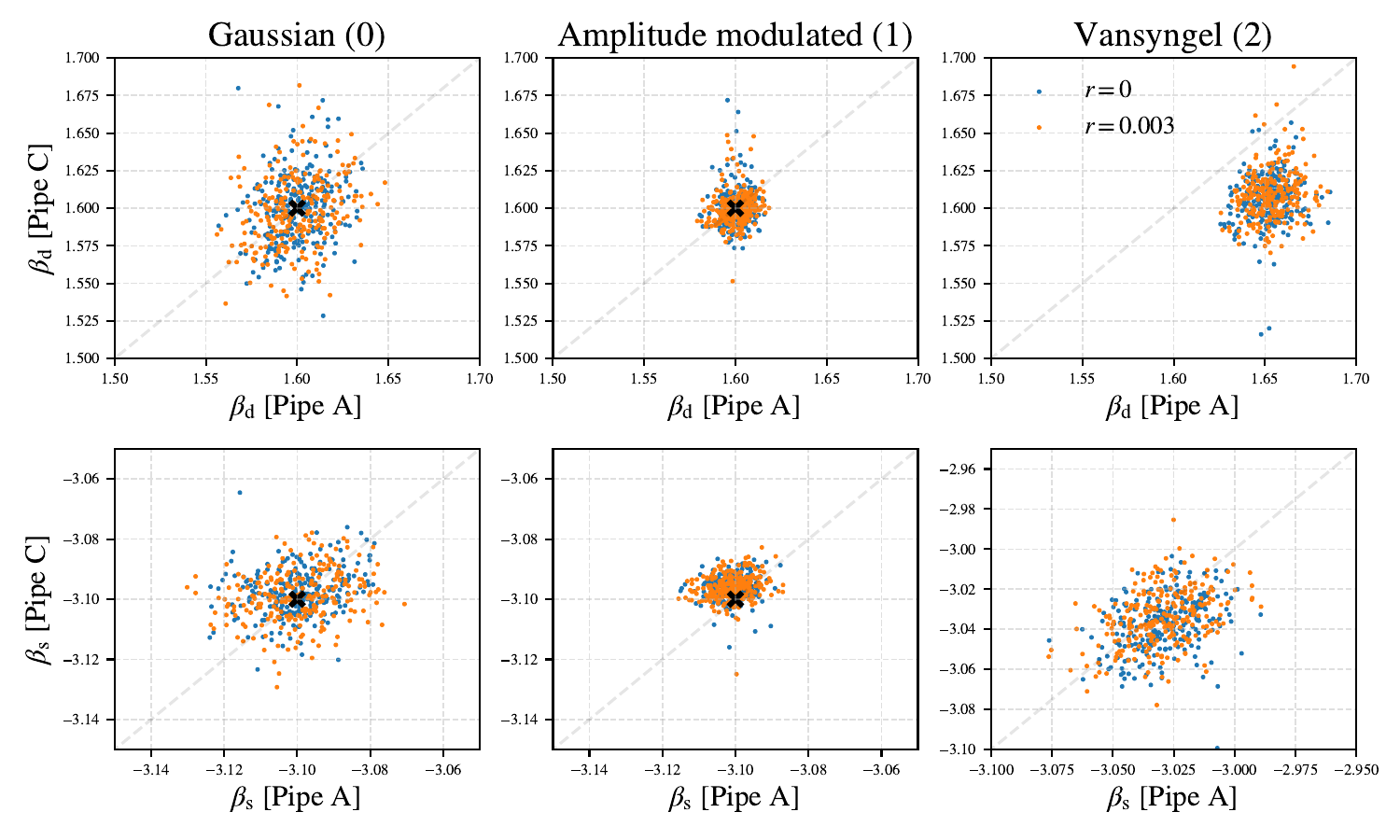}
    \caption{Scatter plots displaying the inferred foreground parameters $\beta_{\rm d}$ (dust spectral index, top row) and $\beta_s$ (synchrotron spectral index, bottom row) from two component-separation methods: map-based maximum likelihood (Pipeline C) and cross-spectral likelihood (Pipeline A). Each column corresponds to a different foreground suite: Gaussian (Model 0), amplitude modulated (Model 1), and Vansyngel (Model 2), from left to right. Points are color-coded to represent different tensor-to-scalar ratio scenarios, with blue representing $r = 0$ and orange representing $r = 0.003$. The black cross in each panel marks the input values used in the simulations, where applicable. In all cases, except for $\beta_{\rm d}$ in Model 2, Pipelines A and C yield consistent dust parameter measurements.}
    \label{fig:scatter_plot_fg}
\end{figure*}

Fig.~\ref{fig:scatter_plot_fg} further explores the recovered spectral indices of dust ($\beta_{\rm d}$) and synchrotron ($\beta_{\rm s}$) by comparing maximum-likelihood estimates from the Pipelines A and C. 
The analysis, repeated for every foreground suite, reveals similar scatter in $\beta_{\rm d}$ and $\beta_{\rm s}$ values across the $r=0$ and $r=0.003$ cases. 
The Gaussian case shows mean recovered $\beta_{\rm d}$ and $\beta_{\rm s}$ consistent with input values, highlighting the robustness of the cleaning methods under standard conditions. 
The amplitude-modulated scenario, with its stronger foreground emission, narrows the variance in estimated spectral indices approximately twofold, enhancing the signal-to-noise ratio. 
Both Pipelines A and C measure an effective, spatially constant $\beta_{\rm d}$ parameter, while the dust SED across the Vansyngel model sky varies. This effective parameter can be different for both pipelines given that they operate in different basis spaces, leading to an offset in the mean recovered value for $\beta_{\rm d}$.\\

The correlation of measured spectral parameters between Pipelines A and C is generally low, with correlation coefficients between 10 and 40\%. 
Moreover both pipelines measure the spectral parameters to high signal-to-noise ratios. Consequently their statistical uncertainty is dominated by noise modes, which do not necessarily overlap between the different pipelines.

\end{document}